\documentclass[]{pasj01}

\usepackage{url}

\begin{document} 
\Received{2017/07/13}
\Accepted{2017/08/21}

\title{Morphological evidence for a past minor merger in the Seyfert galaxy NGC\,1068\thanks{Based on data collected at Subaru Telescope, which is operated by the National Astronomical Observatory of Japan.}}

\author{Ichi \textsc{Tanaka}\altaffilmark{1}%
}
\altaffiltext{1}{Subaru Telescope, National Astronomical Observatory of Japan, 650 North A'ohoku Place, Hilo, Hawaii, 96720, U.S.A. }
\email{ichi@naoj.org}

\author{Masafumi \textsc{Yagi},\altaffilmark{2}}
\altaffiltext{2}{National Astronomical Observatory of Japan, 2-21-1, Osawa, Mitaka, Tokyo, 181-8588, Japan}
\email{YAGI.Masafumi@nao.ac.jp}

\author{Yoshiaki \textsc{Taniguchi}\altaffilmark{3}}
\altaffiltext{3}{The Open University of Japan, 2-11 Wakaba, Mihama-ku, Chiba 261-8586, Japan}
\email{yoshiaki-taniguchi@ouj.ac.jp}

\maketitle

\begin{abstract}
Deep optical imaging with both Hyper Suprime-Cam and Suprime-Cam on the 8.2 m Subaru Telescope reveals a number of outer faint structures around the archetypical Seyfert galaxy NGC\,1068 (M\,77). 
We find three ultra diffuse objects (UDOs) around NGC\,1068. Since these UDOs are located within the projected distance of 45 kpc from the center of NGC\,1068, they appear to be associated with NGC\,1068. 
Hereafter, we call them UDO-SW, UDO-NE, and UDO-SE where UDO = Ultra Diffuse Object, SW = south west, NE = north west, and SE = south east; note that UDO-SE was already found in the SDSS Stripe 82 data.
Among them, both UDO-NE and UDO-SW appear to show a loop or stream structure around the main body of NGC\,1068, providing evidence for the physical connection to NGC\,1068. We consider that UDO-SE may be a tidal dwarf galaxy. 
We also find another UDO-like object that is 2 magnitudes fainter and smaller by a factor of 3 to 5 than those of the three UDOs. This object may belong to a class of low surface brightness galaxy. 
Since this object is located along the line connecting UDO-NE and UDO-SW, it is suggested that this object is related to the past interaction event that formed the loop by UDO-NE and UDO-SW, thus implying the physical connection to NGC\,1068. 
Another newly-discovered feature is an asymmetric outer one-arm structure emanated from the western edge of the outermost disk of NGC\,1068 together with a ripple-like structure at the opposite side. These structures are expected to arise in a late phase of a minor merger according to published numerical simulations of minor mergers. All these lines of evidence show that NGC\,1068 experienced a minor merger several billions years ago. We then discuss the minor-merger driven triggering of nuclear activity in the case of NGC\,1068.
\end{abstract}

\section{Introduction}

\noindent 

NGC\,1068 is one of the archetypical Seyfert galaxies in the nearby Universe \citep{Sey43} and belongs to a class of active galactic nuclei (AGNs). 
Since the majority of Seyfert galaxies are located in the nearby Universe, many multi-wavelength observations have been made to date. Among such Seyfert galaxies, much attention has been paid to NGC\,1068 because this galaxy plays an important role for a unified model of AGNs (\cite{AM85}; see for a review, \cite{UP95}).

Since luminous AGNs like quasars show morphological and dynamical evidence for major mergers between or among galaxies, it has been widely accepted that they are triggered by major mergers \citep{S88,TS98,H08}. On the other hand, most Seyfert galaxies appear to be ordinary-looking spiral galaxies \citep{malkan98}. Indeed, NGC\,1068 is classified as a symmetric spiral galaxy without any morphological disturbance \citep{RC3}. However, taking all available observational properties into account, \citet{Tani99} proposed that the Seyfert activity is triggered by a past merger with a satellite galaxy; i.e., a minor merger (see also \cite{Tani13}, \cite{furu_tani16} and references therein). 

It is noted that morphological evidence for major mergers can be easily obtained because tidal remnants are prominent in most cases and double nuclei can also be seen in some cases. Meanwhile, it has been thought that it is generally difficult to find firm morphological evidence for minor mergers because their dynamical effect on the host galaxy is weak and smeared out during several rotations of the galaxy (e.g., \cite{ji14, Khan12}). Optical images obtained by the Sloan Digital Sky Survey (hereafter SDSS) have been often used to study morphological studies of nearby galaxies because its wide-field sky coverage and homogeneity (e.g., \cite{F07}; \cite{willett13}, and references therein). However, the survey depth of SDSS is not deep enough ($\sim 25$ mag arcsec$^{-2}$) to investigate faint structures in outer parts of many galaxies. Indeed, the Stripe 82 data, 2 magnitude deeper than the original SDSS, reveal such faint outer structure, providing morphological evidence for minor and/or major mergers for a sample of ordinary-looking galaxies \citep{Schawinski10, willett13}. Several faint tidal features have also been found around galaxies in previous studies at 26 to 29 mag arcsec$^{-2}$ level (\cite{martinez-delgado08}, 2010, 2012, 2015; \cite{chonis11, miskolczi11}; \cite{duc2014}, 2015). Therefore, although most Seyfert galaxies show little evidence for minor mergers, it seem worthwhile to re-visit their optical morphology by using much deeper optical images (e.g., \cite{smirnova10}).

Here, we present our discovery of faint outer structures around the main body of NGC\,1068 based on the data obtained with Hyper Suprime-Cam (hereafter HSC; \cite{Miyazaki12}) and Suprime-Cam \citep{Miyazaki2002} on the 8.2 m Subaru Telescope.
The distance toward NGC\,1068, 15.9 Mpc (\cite{KH2013}; we also follow their assumed WMAP 5-yr cosmology), is adopted throughout this paper; $m-M = 31.01$ and $1\arcsec = 0.077$ kpc. 
We use the AB magnitude system unless otherwise noted.
The single Galactic extinction value in $r$ ($A_{r}=0.08$ from the NASA Extragalactic Database; hereafter NED\footnote{$\langle$\url{http://ned.ipac.caltech.edu}$\rangle$}) has been applied to the photometry.

\section{Observations and data analysis}
\subsection{Hyper Suprime-Cam observations}
Our deep $r2$-band (basically the same as the SDSS $r$ filter, and we simply call $r$-band hereafter) HSC data were taken on 2016 December 25 (UTC) by the Queue observing mode (S16B-QF187; Tanaka et al.). The result of the On-Site Quality-Assurance System for HSC (the OSQAH: Furusawa et al. 2017 in prep.) indicates that the transparency condition was photometric, though the seeing condition was as poor as 2 arcsec FWHM at the beginning then improved to $\sim$ 1 arcsec.

Making a good median sky is critically important for the reliable detection of the extremely faint features around galaxies with large apparent size. To achieve this, we used a 5-step HSC-standard dither sequence for the science data acquisition\footnote{$\langle$\url{https://www.naoj.org/Observing/queue/HSC_QUD/node16.html}$\rangle$} but with a fairly big dither size of 720 arcsec radius. The size of the dither circle has more than three times larger than the optical diameter of NGC\,1068 ($D_{\rm maj}\sim7.1$ arcmin.; from NED). The wide field of view of HSC ($\phi=1\degree.5$) makes such big dither possible, and is a great advantage of the HSC for such nearby galaxy science.

The unit exposure was set to 80 seconds. It was chosen to balance between the efficiency and the saturation of the central region. Based on the surface photometry of NGC\,1068 by \citet{SanchezPortal00}, we designed to allow saturation of the very central region ($<10$ arcsec) around the nucleus by the setting, which is later confirmed by the actual data. In total, 2800 sec exposure was gathered with two different position angles; $0^{\circ}$ and $90^{\circ}$.

\begin{figure*}
 \begin{center}
  \includegraphics[width=17.cm]{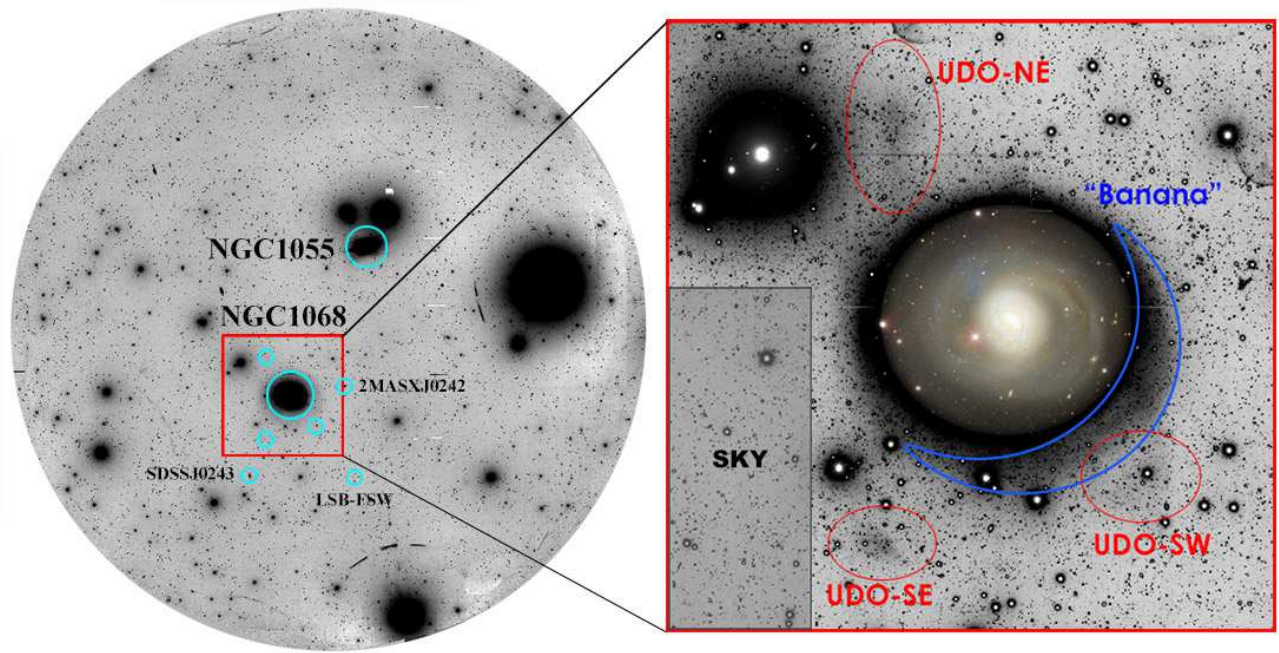} 
 \end{center}
\caption{(\textit{left}) The full 3.14 deg$^{2}$ HSC $r$ band image. The north is up and the east is left. The two member of the NGC\,1068 group (NGC\,1068 and NGC\,1055), as well as the spectroscopically-confirmed member dwarf galaxies (
SDSS\,J024310.38-001543.2 and 2MASX\,J02420036+0000531 labeled as SDSS\,J0243 and 2MASX\,J0242, respectively) are shown as cyan circles. 
 Another newly-found compact diffuse dwarf ``LSB-FSW'' is also shown. 
(\textit{right}) The $23\arcmin \times 23\arcmin$ cutout around NGC1068. The three newly-discovered UDOs around NGC\,1068 (see section 3) are shown with the ovals with labels.
The color image of NGC\,1068 from the SDSS (courtesy of Michael R. Blanton) is overlaid on the HSC data to show the the detail of the inner structure. 
A faint outer one-arm structure is shown as blue crescent labeled as ``Banana'' (see section 3 as well as figures \ref{fig3} and \ref{fig4a}). The gray region labeled "SKY" is the area used for the depth measurement (see section 2.3).}\label{fig1}
\end{figure*}

Data reduction is performed using the HSC Pipeline (HscPipe ver. 4.0.5) distributed by the instrument team\footnote{see $\langle$\url{http://hsc.mtk.nao.ac.jp/pipedoc_e/}$\rangle$}. The algorithm of the pipeline is described in \citet{bosch17} in detail. We have applied a special sky pattern subtraction routine (Koike, M. et al. 2017, private comm.\footnote{$\langle$\url{http://hsc.mtk.nao.ac.jp/pipedoc_e/e_tips/skysub.html}$\rangle$}) that works better for the data with large objects than the HscPipe standard procedure does. Firstly the program makes a median sky image from all exposures in each CCD after masking visible objects and normalizing the sky level. The median sky image is scaled to the sky values on each exposure data for each CCD and then subtracted. Since, at this stage, we still see some large-scale residual patterns, we have to remove these patterns in order to investigate very faint structures; i.e., we need much better object masks for apparently large objects. Therefore, the special routine tries to fit and then subtract the residual sky pattern by using the 3rd-order polynomials. 
Finally, the polynomial fitting for the sky in finer scale is applied by using the new more complete object masks for sky subtraction.  Our observing strategy just matches to the sky-handling method used by this program. The final co-added image by the procedure is shown in figure~\ref{fig1}. 

The photometric and astrometric calibrations were done by the HSC Pipeline using the DR1 catalog from the Pan-STARRS1 \citep{chambers16}. 
The photometric and the astrometric accuracy measured by the pipeline (rms) is $\sim0.018$ mag and $\sim0\arcsec.03$, respectively
\footnote{In this study, we did not take the aperture correction adopted in the Pipeline into account, which makes a slight offset in the surface brightness according to the point spread function (Furusawa, H., priv. comm.). The possible offset would be $<0.1$ mag at most, and does not affect our results.
}.　We discuss about the depth of the HSC image in section 2.3 below.

\subsection{Archived Suprime-Cam data}
In addition to our own HSC imaging, we have also searched for the Subaru data archive (SMOKA: \cite{Baba2002}) to see if there are any other deep imaging data for NGC\,1068 available. Since our main purpose is to find evidence
for very faint structures around NGC 1068, any independent observations are highly useful to confirm their presence if any. We then have found the $R_\mathrm{c}$-band images of NGC\,1068. The observations were made on 2014 December 17 (UTC), with 15 of 150-second exposures and 15 of 20-second exposures (in total 2550 seconds). We note that the data have become public after we submitted the HSC queue proposal for the current study.

The data reduction is the same as \citet{yagi2017} except for the 
background subtraction. As our targets in this study are widely
extended objects, we first connect 10 CCDs of the same exposure and
estimate a constant background of the exposure across the CCDs and
subtract the value. After median coadding, we subtract the background
by estimating in a mesh size of 2000 pixels (6.73 arcmin) square.

We use astrometry.net \citep{Lang2010} for the astrometric calibration.
Flux is calibrated using SDSS DR12 stars 
\citep{SDSS-DR12}
with a color conversion to the Suprime-Cam system. 
The color conversion procedure is given in \citet{yagi2013}, 
and the coefficients of the color conversion are given in \citet{Yoshida2016}.
About 2700 stars in $18<r<20$ are used for the flux zero-point 
estimation. The rms estimated from the median of absolute deviation
(MAD) is 0.04 mag. We also checked that the data were taken under good condition from the investigation of the photometric zero point of each exposure. 

\subsection{The depth of the data}
Here we discuss about the depth of the data we use. We employ the random-aperture photometry test: we measure the fluctuation of the flux inside a fixed aperture radius for a randomly-selected blank sky area.

We firstly cut out a part of the image near NGC\,1068 where no bright stars or galaxies exist ($5\arcmin.6 \times 12\arcmin.6$: the area labeled as "SKY" in the right panel of figure~\ref{fig1}). After removing the residual large-scale sky pattern with 3rd-order polynomials by the IRAF \texttt{imsurfit}, we run the SExtractor (version 2.9.1; \cite{bertin}) for the image and try to detect all the visible objects as much as possible. Using the "OBJECT" frame generated by the software, we create the object mask image. Then we generate random sky positions in the SKY image. Every time each random position is generated, we check if it is away from the areas covered by the object masks. We set the minimum acceptance distance from the masked areas to $2\arcsec.0$. We should note that larger than this acceptance separation makes the blank area for measurement extremely limited to some particular under-dense regions, as our image is very deep and most of the sky is occupied by faint objects. After generating 1000 good empty sky positions, we put aperture with the radius of $0\arcsec.5$ to $2\arcsec.5$ at each position, and measure the counts inside the apertures by the IRAF \texttt{phot}. Since the pixel scales are $0\arcsec.168$ pixel$^{-1}$ (HSC) and $0\arcsec.202$ pixel$^{-1}$ (Suprime-Cam), the adopted radii correspond to 3.0--14.9 pixels (HSC) and 2.5--12.4 pixels (Suprime-Cam), respectively. We then measure the standard deviation of the counts after applying a 3-$\sigma$ clipping.
Final results are derived by converting the counts in each aperture to mag arcsec$^{-2}$ unit. We do the same procedure to both the HSC and the Suprime-Cam data, using the same sky region defined in the right panel of figure~\ref{fig1}.

\begin{figure}
 \begin{center}
  \includegraphics[clip,width=6cm,angle=-90]{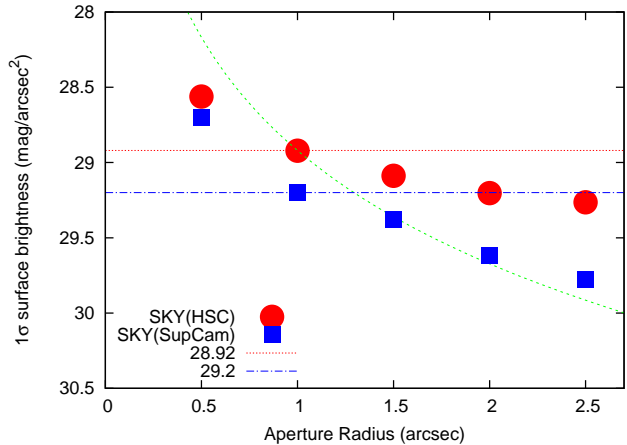} 
 \end{center}
\caption{The 1-$\sigma$ sky fluctuation of the background of our data used. The measured 1-$\sigma$ sky fluctuation in the given aperture radius is converted to the surface brightness unit. The green curve is the theoretical curve for the pure Poisson noise data in arbitrary scale.
}\label{fig8}
\end{figure}

In figure~\ref{fig8} we show the 1-$\sigma$ level of the sky fluctuation as a function of the aperture radius (in $r$ band for HSC, while $R_\mathrm{c}$-band in Suprime-Cam). The depth at the $1\arcsec$-radius aperture for the HSC data is 28.9 mag arcsec$^{-2}$, while it is 29.2 mag arcsec$^{-2}$ for Suprime-Cam data. If we assume the color of the typical galaxy ($r-R_\mathrm{c} \sim 0.05$ to 0.1, based on the filter functions available on the Subaru website\footnote{$\langle$\url{https://www.naoj.org/}$\rangle$} and the galaxy SED models by \cite{cww80}), the Suprime-Cam data is about 0.3 mag deeper than the HSC data, though the exposure time is similar. There are a few possible reasons for this. Firstly, the seeing was quite poor ($\sim 2\arcsec$) at the first block of the data acquisition of HSC. Slower growth of the depth for HSC compared to the Suprime-Cam data points suggests that the blank sky region is actually affected by faint objects below the detection threshold. Secondly, our HSC data were taken in late 2016, just around the period when a significant degradation of the reflectivity of the primary mirror was found\footnote{This was given in the telescope status report during the Subaru Users Meeting 2016; see $\langle$\url{https://www.subarutelescope.org/Science/SubaruUM/SubaruUM2016/}$\rangle$.}.
Thirdly, a part of (5 frames or 15\%) our HSC data was taken at the elevation somewhat lower (40 to 43 degrees), though the other 85\% was take at the similar elevation (60 to 69 degrees) to the Suprime-Cam data (66 to 70 degrees). The effect of the atmospheric extinction for HSC data in this study, however, should be negligibly small, as we confirmed that our data were taken under the photometric condition via our on-site quality assessment tool as well as several meteorology data of the Maunakea Observatories.

\section{Results}

\subsection{A brief summary of our findings}

The main purpose of our study is to see if there are any signatures of past interaction/mergers which may have led the current AGN activity observed in NGC\,1068. 
As the dynamical timescale of the outer region of galaxies could extend as long as several Gyrs (e.g., \cite{ji14, martinez-delgado08}),
 we may expect to see, if any, remnants of such a past dynamical perturbation event. 

Based on our deep HSC image, we have identified six objects summarized in table~\ref{tbl3}. Among them, three are very faint objects, being similar to ultra diffuse galaxies (UDGs) found in galaxy cluster environments (e.g., \cite{vDm15,koda15,yagi2016}). Since all these are close to the main body of NGC\,1068 (their projected distance from NGC\,1068 are less than 45 kpc; see table~\ref{tbl2}), they are considered to be associated with NGC\,1068. Their central surface brightness is fainter than 26 mag arcsec$^{-2}$ ($r$ band) and the effective radius of these galaxies are larger than 1.5 kpc. Assuming that $g-r$ color is not extremely red ($g-r<1.5$: note that the typical $g-r$ color of galaxies is $g-r<0.77$; \cite{fukugita95}), we find that these three objects satisfy the criteria for the UDGs \citep{vDm15}.
Taking account of this similarity, we call them UDO-SW, UDO-NE, and UDO-SE where UDO = Ultra Diffuse Object, SW = southwest, NE = northwest, and SE = southeast; note that UDO-SE was already found in the SDSS Stripe 82 data \citep{bakos12}. These three UDOs are shown in the right panel of figure~\ref{fig1}.

For a cross-check, we examine the archived Suprime-Cam $R_\mathrm{c}$-band data, and securely identify these three UDOs. We further examine the SDSS FITS data \citep{SDSS-DR7}\footnote{We used the data from both DR7 (SDSS-II final release $\langle$\url{http://cas.sdss.org/dr7/en/}$\rangle$) and the DR12 (SDSS-III final release $\langle$\url{https://dr12.sdss.org/}$\rangle$), because the background pattern of the UDO-SE field in the DR7 is nicer than the DR12, while the DR12 image is deeper.}. However, we find that they are too faint to be securely detected, though we see a hint of UDOs after heavy smoothing. Therefore, we conclude that these UDOs are real objects (see Appendix for details). 
It is intriguing to note that both UDO-NE and UDO-SW appear to show a loop or stream structure around the main body of NGC\,1068, providing evidence for the physical connection to NGC\,1068 (see section 4.1).

We also find another UDO-like object that is 2 magnitudes fainter and smaller by a factor of 2 than those of the three UDOs. This object may belong to a class of low surface brightness galaxy. We call this object, LSB-FSW (F means `far').
Note that this object is out of the field of views in the Suprime-Cam $R_\mathrm{c}$-band data. Although it seems uncertain if this object is physically associated with NGC\,1068, we assume that this object is also a member of M\,77 group and a possible companion galaxy. It is interesting to note that this object is located at the line connecting both UDO-NE and UDO-SW.
Therefore, we suggest that this object may be a relic of the formation of the loop structure.

The other two galaxies are identified as SDSS\,J024310.38-001543.2 (hereafter SDSS\,J0243)\footnote{We note that there are three IDs on the position of the object in the NED (others are SDSS\,J024310.55-001546.1 and SDSS\,J024310.39-001543.8). According to the SDSS Sky Server, the coordinate of the referred object better refers to the position of the object.} and 2MASX\,J02420036+0000531 (hereafter 2MASX\,J0242). They are apparently normal dwarf galaxies. Since their redshifts, 0.00370 and 0.00362 (SDSS DR13), respectively, are very close to that of NGC\,1068 ($z=0.003793$, \cite{RC3}), they are considered as companion galaxies of NGC\,1068. The spectroscopic information of the two dwarf galaxies are given by SDSS DR13. The spectrum of SDSS\,J0243 shows strong H$\delta$ absorption (Lick H$\delta_\mathrm{A}$ index $>$ 5.0 \AA) without H$\alpha$ emission. The spectral features suggest that the galaxy is a post-starburst galaxy (e.g., \cite{Goto2005}). The spectrum of 2MASX\,J0242 shows strong emission lines. The [N\emissiontype{II}]/H$\alpha$ and [O\emissiontype{III}]/H$\beta$ ratios indicate that the galaxy is a typical star-forming galaxy \citep{Baldwin1981,Kewley2001,Kauffmann2003}.

\begin{table*}[!t]
 \tbl{Summary of the identified structures around NGC\,1068}{
 \begin{tabular}{cccc}
     \hline
     Name & Note 1 & Note 2 & Redshift ($z$) \\
     \hline
     UDO-NE & Newly discovered & A part of loop?  & --- \\
     UDO-SW & Newly discovered & A part of loop?  & --- \\
     UDO-SE & \citet{bakos12} & ---  & --- \\
     LSB-FSW & Newly discovered &  --- & --- \\
     SDSS\,J0243 & SDSS DR13 & Star forming & 0.00370 \\
     2MASX\,J0242 & SDSS DR13 & Post starburst & 0.00362 \\
     \hline
     Banana & Newly discovered & Asymmetric one arm emanated from the western edge  &  --- \\
     Ripple & Newly discovered & Eastern side of the outer disk & --- \\
     \hline
   \end{tabular}}\label{tbl3}
   \begin{tabnote}
Note: the "ripple" and "banana" structures we claim discovery in this work can actually be visible in a few very high-quality public images posted on websites. A good examples is the one by Adam Block at the Mount Lemmon Observatory $\langle$\url{http://www.messier-objects.com/messier-77-cetus-a/}$\rangle$.
\end{tabnote}

\end{table*}

In addition to the above six objects near NGC\,1068,
we find an asymmetric outer one-arm structure in the main body of NGC\,1068, emanated from the western edge of the disk of NGC\,1068 
through south to east, together with a ripple-like structure at the opposite side. Since these features are not expected to be present in isolated spiral galaxies, they may provide another line of evidence for a past minor merger.

\subsection{Observational properties of the UDOs and the other companion galaxies}

As described above, the three UDOs appear to be associated with NGC\,1068 although there is no redshift information. In order to understand what they are, it is important to investigate their structural properties in detail. It is, however, reminded that there are some difficulties in doing such analysis because of the followings.
First, they are very faint. Second, since they are close to NGC\,1068 itself, their light profiles may be affected by the outskirt
of NGC\,1068 light. Third, there are some bright stars near UDO-NE and thus
the ghost by the HSC optics overlaps with it at the faintest level. Despite these difficulties, we here try to fit their light profile carefully.

We use the model galaxy fitting code GALFIT version 3.0.5 \citep{peng02, peng10} to derive the physical parameters. We choose the single S{\'e}rsic profile \citep{sersic} for fit. In order to have the reliable result from the GALFIT software, the careful sky subtraction is critically important. To achieve this, we first make the catalog of all objects in the cut-out images around the dwarf galaxies on and around the faint objects using the SExtractor (version 2.9.1: \cite{bertin}). We then use the ``OBJECT'' frame generated by SExtractor for the object mask, after expanding the object regions by 2--5 pixels. The object mask made by the process is also used during the GALFIT process later. We further replace the areas with the object mask with the average sky values sampled about $100 \time 100$ pixels around each area. However, the object mask by the process does not include the UDO because it is so diffuse. We then replace the area around the UDO with a sky surface using the IRAF task \texttt{imedit}. Here we pay a special attention so that the profile of the interpolated sky surface is smoothly connected to the area outside. The object-eliminated frame by the process is then used for the sky pattern fitting by the two-dimensional spline function (3 terms) by the IRAF task \texttt{imsurfit}. The fitted sky surface is finally subtracted from the original cut-out images and processed by GALFIT. As the sky region is actually the outer profile of NGC\,1068 and/or the halo of the nearby bright stars for UDO-NE and UDO-SW, this is the critical process for GALFIT to work properly (see below). 

The basic properties measured or calculated by GALFIT for the six identified objects are summarized in table~\ref{tbl1}. The size of the detected objects is measured by using the isophote at 28.7 mag arcsec$^{-2}$ (which corresponds to $\sim 1\sigma$ surface brightness level with a $1\arcsec$-diameter aperture) after applying the $5\times5$ pix$^{2}$ boxcar smoothing to the data to suppress the noise. The residual image after subtracting the best-fit model for each object is shown in the second row of figure~\ref{fig3}. Because the objects are widely extended, we do not use automatic estimation of the sky level by GALFIT but explicitly give 0 for the sky level after carefully subtracting the sky component.

{The estimate of the physical parameters (S{\'e}rsic index, $r_{e}$ etc.) by GALFIT tends to be quite sensitive to the error of the sky background level (e.g., \cite{haussler07}). As the errors estimated by GALFIT do not work to estimate the uncertainty introduced by the error of the sky background level, we simulate how the output of parameters by GALFIT changes with the error in the sky background level. We conservatively set the range of the error in the background level as $\pm 0.1 \sigma$ of the sky noise. The range of the change on each parameter is given in table~\ref{tbl1}. 

\begin{figure*}
 \begin{center}
  \includegraphics[width=16cm]{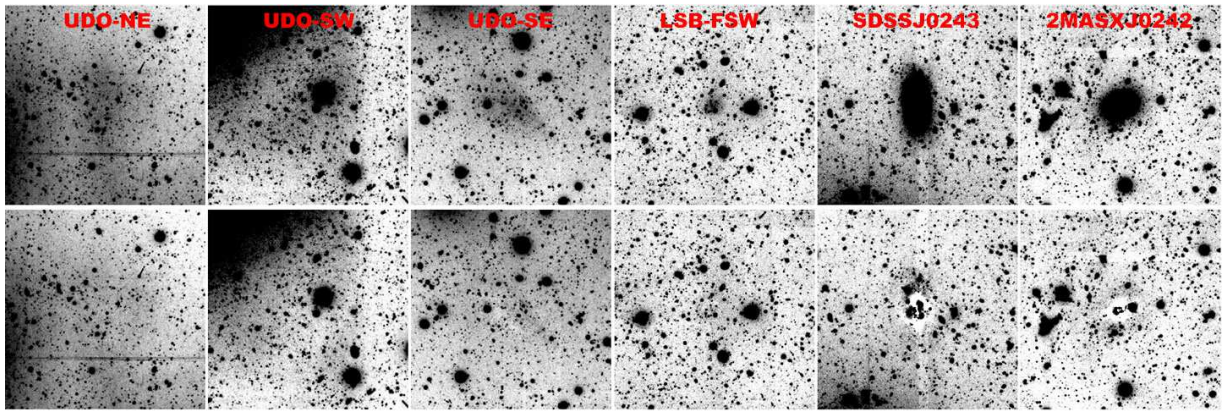} 
 \end{center}
\caption{The newly-discovered ultra-diffuse objects (from left to right, UDO-NE, UDO-SW, UDO-SE), as well as a candidate faint companion dwarf (LSB-FSW), and the known two bright companion dwarf galaxies (SDSS\,J024310.38-001543.2 and 2MASX\,J02420036+0000531) are shown at the top. The images in the second row are after subtracting the best-fit model galaxy by GALFIT. Each cutout has $4\arcmin.48$ on a side.}\label{fig3}
\end{figure*}

\begin{table*}[t]
  \tbl{The structural properties of the six identified objects around NGC\,1068}{
  \begin{tabular}{ccccccccccc}
      \hline
      Name & R.A. & Dec. & distance & $r$ mag & $\mu_{0}$ & Size  & S{\'e}rsic $n$  & $r_\mathrm{e}$ & $b/a$ & $\langle\mu(r_\mathrm{e})\rangle$ \\
           & J2000.0 & J2000.0 & arcmin. &  & mag arcsec$^{-2}$ & arcmin. &   & arcsec. & & mag arcsec$^{-2}$ \\
      \hline
      UDO-NE      & \timeform{02h42m57s.65} & $+$\timeform{00D06'18".5} &  8.58 & 17.3 $\pm0.4$  & 26.97 $\pm0.2$  & $ 2.7 \times 1.2 $ & 0.40\,$^{0.31}_{0.66}$ & 58\,$^{48}_{77}$ & 0.44 & 28.1\,$^{27.7}_{28.8}$ \\
      UDO-SW      & \timeform{02h42m21s.57} & $-$\timeform{00D06'45".2} &  7.52 & 17.5 $\pm0.5$  & 27.02 $\pm0.3$  & $ 3.3 \times 1.4 $ & 0.39\,$^{0.26}_{0.83}$ & 70\,$^{60}_{106}$ & 0.53 & 28.7\,$^{28.4}_{29.6}$  \\
      UDO-SE      & \timeform{02h43m00s.48} & $-$\timeform{00D09'00".1} &  9.48 & 17.6 $\pm0.1$  & 26.40 $\pm0.05$ & $ 1.8 \times 1.0 $ & 0.38\,$^{0.31}_{0.89}$ & 38\,$^{34}_{54}$ & 0.59 & 27.5\,$^{27.3}_{28.3}$  \\
      LSB-FSW     & \timeform{02h41m53s.51} & $-$\timeform{00D15'49".3} & 19.11 & 19.4 $\pm0.1$  & 26.19 $\pm0.03$  & $ 0.7 \times 0.6 $ & 0.45\,$^{0.34}_{0.84}$ & 12.2\,$^{10.8}_{17.5}$ & 0.91 & 26.9\,$^{26.6}_{27.8}$  \\
      SDSSJ0243  & \timeform{02h43m10s.46} & $-$\timeform{00D15'44".4} & 16.69 & 15.40 $\pm0.05$ & 21.92 $\pm0.02$ & $ 2.0 \times 1.1 $ & 1.12\,$^{1.08}_{1.18}$ & 17.4\,$^{17.0}_{18.0}$ & 0.54 & 23.54\,$^{23.50}_{23.60}$  \\
      2MASXJ0242 & \timeform{02h42m00s.38} & $+$\timeform{00D00'52".3} & 10.22 & 15.46 $\pm0.05$ & 20.24 $\pm0.02$ & $ 1.4 \times 1.2 $ & 2.02\,$^{1.94}_{2.12}$ & 8.9\,$^{8.7}_{9.2}$  & 0.67 & 22.22\,$^{22.17}_{22.27}$  \\
      \hline
    \end{tabular}}\label{tbl1}
\begin{tabnote}
Notes: SDSS\,J0243 and 2MASX\,J0242 refer to SDSS\,J024310.38-001543.2 and 2MASX\,J02420036+0000531, respectively. The distance refers to the projected distance from the center of NGC\,1068. The $r$ mag and the central surface brightness $\mu_{0}$ are the observed value (in AB). The size of the detected objects is measured in the 28.7 mag arcsec$^{-2}$ isophote. The effective surface brightness ($\langle\mu(r_\mathrm{e})\rangle$) is the average surface brightness within the radius of $r_\mathrm{e}$. The range of uncertainties of the physical parameters by GALFIT is estimated by artificially changing the sky level by $\pm 0.1 \sigma$. The superscript and subscript numbers refer to the value for the case where the sky estimate is wrong by $+0.1\sigma$ and $-0.1\sigma$, respectively.
\end{tabnote}
\end{table*}

\subsection{Distorted outer structures of NGC\,1068}

It is known that the disk structure of NGC\,1068 at high surface-brightness level is quite symmetric. Its two spiral arms with the outer ring structure do not show any non-axisymmetric signature. At a first glance, one may consider that NGC\,1068 is a spiral galaxy without any sign of galaxy interaction \citep{holwerda14}. However, any signatures of past dynamical disturbance is often found at faint outer envelope regions of galaxies (e.g., \cite{martinez-delgado10}). 
Therefore, in order to see such signatures, it is necessary to investigate NGC\,1068 at very faint brightness levels.

As shown in the right panel of figure\ref{fig1},
there is an evident one-arm structure at the south-west of the disk: we call this ``Banana''. In order to see this structure more clearly, we here employ the unsharp-masking technique to enhance the contrast. 

To do this, we first detect all the foreground/background (compact) objects in the image using the SExtractor and eliminate them using the adjacent average count around the detected objects (``cleaned image''). For more complete detection of objects on the disk part of NGC\,1068, we first apply large median filter (21 pix by 21 pix) to the data and subtract it from the original. This helps detection of compact source on significant part of the disk. After interpolation is done, we then smoothed it using the Gaussian kernel ($\sigma=50$ pixels) after 5 by 5 block-averaging the data to reduce the image size. In order to avoid having too strong fake negative features, we multiply 0.7 to the smoothed image. Then it is subtracted from the $5 \times 5$ block-averaged version of the cleaned image.

The result is shown in figure~\ref{fig4}. It is clearly shown that there is a clear dip between the Banana structure and the NGC\,1068 main disk. 
Since this dip is actually seen in the image even before the contrast enhancement, this is not due to the artificial effect of the unsharp-masking technique. In figure~\ref{fig4a} we also show the radial surface brightness profile of a part of the NGC\,1068 disk (before unsharp-masking) that includes the Banana region. A narrow rectangular region (35-pixel width) from the center of NGC\,1068 toward southwest direction (position angle $=-135^{\circ}$) is cut out and used for this analysis. The exponential disk is well fit at the high surface-brightness area of the disk (from $200\arcsec$ to $280\arcsec$ area in red), and with a clear discontinuity (labeled as ``dip'') the flux excess at the Banana region is clearly shown.

Another notable feature seen in the unsharp-masked image is a ripple-like structure at the opposite side of the one-arm Banana structure (it is labeled as ``Ripple'' in figure~\ref{fig4}). 
Since they appear at opposite positions, one may consider that they are not physically related.
However, such a pair of unusual structures can be seen in minor merger simulations that reproduce the ring-like features discovered 
in our Milky Way \citep{kazantzidis08, purcell11}. 
This reinforces that NGC\,1068 experienced a past minor merger in its life.
We will discuss this issue more in the next section.

\begin{figure}
 \begin{center}
  \includegraphics[width=7.5cm]{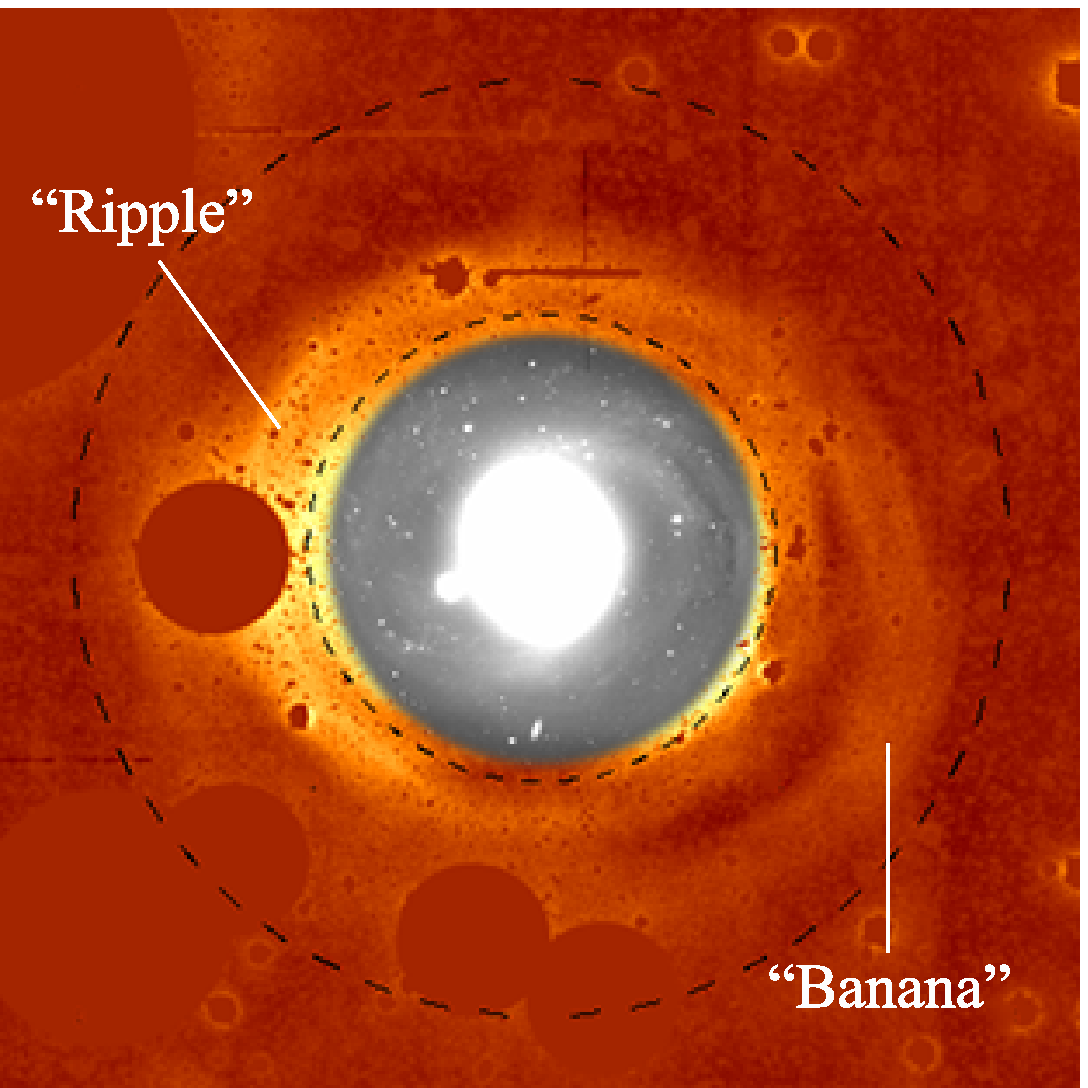}
 \end{center}
 \caption{The detail of the "Banana" structure and the location of the ``Ripple''. The unsharp-masking contrast enhance method is applied to the HSC data after eliminating the foreground/background objects interpolated by the adjacent sky. The dashed small and large circles show the radius of $3\arcmin.4$ and $6\arcmin.8$ from the center, respectively. We also show the original HSC image of NGC\,1068 inside the small circle with gray scale in shallow contrast.
}\label{fig4}
 \begin{center}
 \includegraphics[width=8cm]{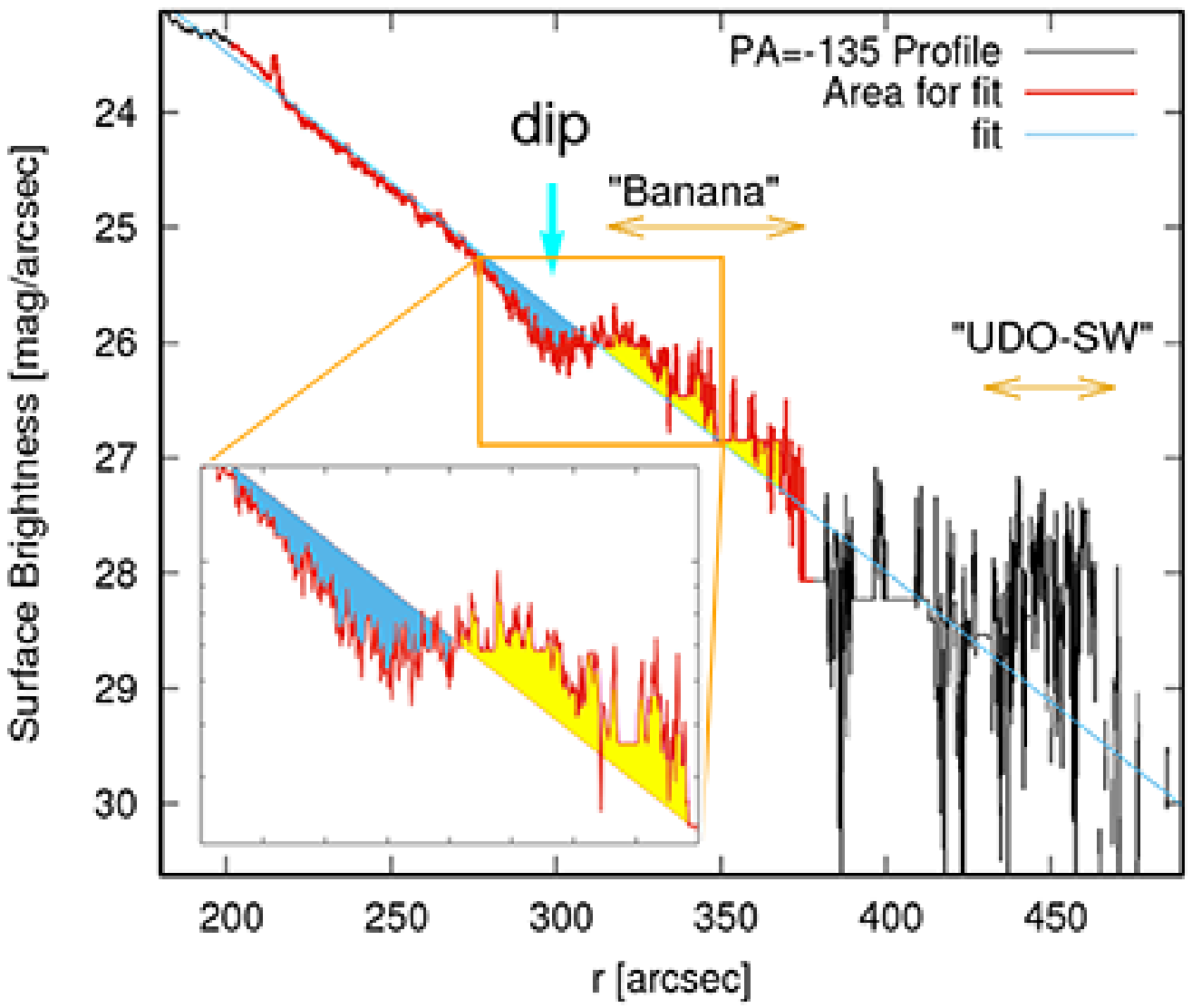}
 \end{center}
 \caption{Radial surface brightness profile toward the southwest direction ($PA = -135$ deg) 
as a function of the distance from the center of NGC\,1068. The black line shows a series of the median count in 35 pixels. All foreground/background objects in the region are replaced to the adjacent counts before radial profile is made.
The orange line shows the fit of the exponential disk profile,
and the fitted range is shown as red points. The inset is the zoomed view of the dip \& the Banana areas. The dip and the Banana region are evidently recognized. We note that the small excess flux at $r\sim200\arcsec$ is the edge of the outer ring.
}
\label{fig4a}
\end{figure}

\section{Discussion}

\subsection{The nature of the NGC\,1068 UDOs}

The three UDOs all have very small S{\'e}rsic index ($n$), while they have similarly large effective radius as the cluster UDGs \citep{vDm15,koda15,yagi2016}.
This is a significant difference from the UDGs found in cluster environment. 
We here compare some physical parameters of our NGC\,1068 UDOs with those of other similar objects from the literature. To do this, the observed values in table~\ref{tbl1} are converted to the physical values in table~\ref{tbl2}, adopting the distance of 15.9 Mpc and assumed all the dwarfs listed are at the same distance.

In figure~\ref{fig7}, we show the following two diagrams; the absolute magnitude versus effective radius (upper panel), and the S{\'e}rsic index versus $\langle\mu(r_\mathrm{e})\rangle$, the average surface brightness within the effective radius (lower panel). In order to compare the properties of both the NGC\,1068 UDOs and the other companions, we also plot the data for the normal dwarf galaxies in the Virgo \citep{gavazzi05} and the Coma \citep{mobasher01, komiyama02} clusters, the UDGs and their analogs in the Coma cluster from \citet{yagi2016} as well as their UDG subsample originally detected in \textit{g}-band by \citet{vDm15}. For the \citet{yagi2016} sample, we use the values from the 1-component S{\'e}rsic fit given in their table~4. Here we convert our photometry values and values from the literature to $R_\mathrm{c}$ in the AB system ($R_\mathrm{c,AB}-R_\mathrm{c,Vega}=0.22$) using the color of the elliptical galaxies from the \citet{cww80} model SEDs calculated by us ($B_\mathrm{Vega}-R_\mathrm{c, Vega}=1.51$ and $r'_\mathrm{AB}-R_\mathrm{c, AB}=0.07$). For the $m-M$ conversion to the Coma and the Virgo clusters, we use the values 35.04 and 30.70 (from NED; here the WMAP 5-yr cosmology is adopted), respectively.

The upper panel of figure~\ref{fig7} shows that our NGC\,1068 UDOs have similarly large size to the UDGs discovered by \citet{vDm15} and \citet{yagi2016}, while they are roughly 1--2 mag fainter. On the other hand, the three companions (2MASX\,J0242, SDSS\,J0243 and LSB-FSW) are located in the sequence of the `normal' dwarfs with a small effective radius. This demonstrates that the NGC\,1068 UDOs are structurally different populations from these `normal' dwarf galaxies.

\begin{table}[t]
  \tbl{The physical properties of the objects in table 2.
  }{
  \begin{tabular}{rccccc}
      \hline
      Name & & distance & $M(r)$ & Size & $r_\mathrm{e}$ \\
           & & kpc     & mag    & kpc &  kpc   \\
      \hline
      UDO-NE & $\cdots\cdots$ & 39.70 & $-13.7$ & $12.7 \times 5.3$ & 4.5\,$^{3.7}_{5.9}$ \\
      UDO-SW & $\cdots\cdots$ & 34.77 & $-13.5$ & $15.3 \times 6.2$ & 5.4\,$^{4.6}_{8.2}$ \\
      UDO-SE & $\cdots\cdots$ & 43.86 & $-13.4$ & $8.3 \times 4.6$ & 2.9\,$^{2.7}_{4.2}$ \\
      LSB-FSW & $\cdots\cdots$ & 88.36 & $-11.6$ & $3.0 \times 2.8$ & 1.0\,$^{0.9}_{1.5}$ \\
      SDSS\,J0243 & $\cdots\cdots$ & 77.19 & $-15.61$ & $9.0 \times 5.1$ & 1.31\,$^{1.28}_{1.34}$ \\
      2MASX\,J0242 & $\cdots\cdots$ & 47.28 & $-15.55$ & $6.6 \times 5.3$ & 0.69\,$^{0.68}_{0.71}$ \\
      \hline
    \end{tabular}}\label{tbl2}
\begin{tabnote}
Notes: see notes in table~\ref{tbl1} for the object names and the uncertainty range in $r_{e}$. We use the distance of 15.9 Mpc for all the objects (see text).
\end{tabnote}
\end{table}

The extreme nature of the NGC\,1068 UDOs is demonstrated in the lower panel of figure~\ref{fig7}. 
They are distributed at the smallest extreme of the S{\'e}rsic index and the faintest end of the average surface brightness\footnote{For reference, we mention that 
 4 UDGs around NGC\,5485 reported by \citet{merritt16} have comparable S{\'e}rsic indices (0.3--0.7) to those of the UDOs in this study.},
while the two companions lie in the sequence of a larger S{\'e}rsic index for brighter surface brightness, made by the Virgo normal dwarfs \citep{graham03}. Here note that we use the distance of 15.9 Mpc for NGC\,1068 from its recession velocity \citep{KH2013}. If we take a shorter distance of $10.3\pm3$ Mpc from the measurements by Tully-Fisher relation (from NED), the absolute magnitudes of the UDOs are estimated as $\sim 1$ magnitude fainter and the scale length become 64\% smaller. 
Moreover, the parameters used in the lower panel, S{\'e}rsic $n$ and 
$\langle\mu(r_\mathrm{e})\rangle$, are unaffected by the change of the distance.
Thus, the extreme nature of the UDOs presented here in figure~\ref{fig7} basically does not change.
It is also interesting to note that LSB-FSW also shares the same property as those of the NGC\,1068 UDOs, while its physical size is 3 to 5 times smaller.
The small S{\'e}rsic index indicates a flat-top stellar distribution \citep{grahamdriver05}.
Such a less centrally-concentrated stellar distribution implies that the
UDOs around NGC\,1068 may have a shallower gravitational potential and thus
lesser content of dark matter than the UDGs found in clusters
\citep{vDm15b,vDm16}.

 The observed extreme nature of the NGC\,1068 UDOs suggests that their origin is different from that of cluster UDGs.
 If the three UDOs are remnants of a past minor merger, one possibility is that they are tidally induced dwarf galaxies (`tidal dwarf' galaxy; \cite{BH92}).
 In fact, UDO-SE appears to be located at the end of the Banana structure.
 On the other hand, as noted before, both UDO-NE and UDO-SW may consist of a loop (or stream) structure around NGC\,1068 because both UDO-NE and UDO-SW show a sign of connection to NGC\,1068. In order to see their morphological property in more detail, we make contrast-enhanced images of these UDOs. The results are shown in figure~\ref{fig5}. Here we use the Suprime-Cam data for UDO-NE while the HSC data for UDO-SW taking account of their relative image quality.
 
If this loop interpretation is correct, its physical size is 74.6 kpc in diameter, with the projected ellipticity ($b/a$) of 0.264. If the loop has the circular shape, the viewing angle toward the loop is estimated as 15.3 degrees.
 Assuming that the dynamical mass of NGC\,1068 within the 74.6 kpc sphere is $3\times10^{11} M_{\solar}$ (\cite{dickel78}, converted to the cosmology we adopted here), the size of the loop indicates the orbital period of 1.2 Gyr. In order to turn the tidal debris into the loop, it may require several orbital timescale as suggested by \citet{martinez-delgado08}. Thus, the first encounter that made the loop could be more than 5 Gyr ago. The core (or nucleus) of the satellite might have already merged into the NGC\,1068 due to the dynamical friction during the interaction with the disk.

\begin{figure}[t]
\begin{center}
\includegraphics[clip,width=6cm,angle=-90]{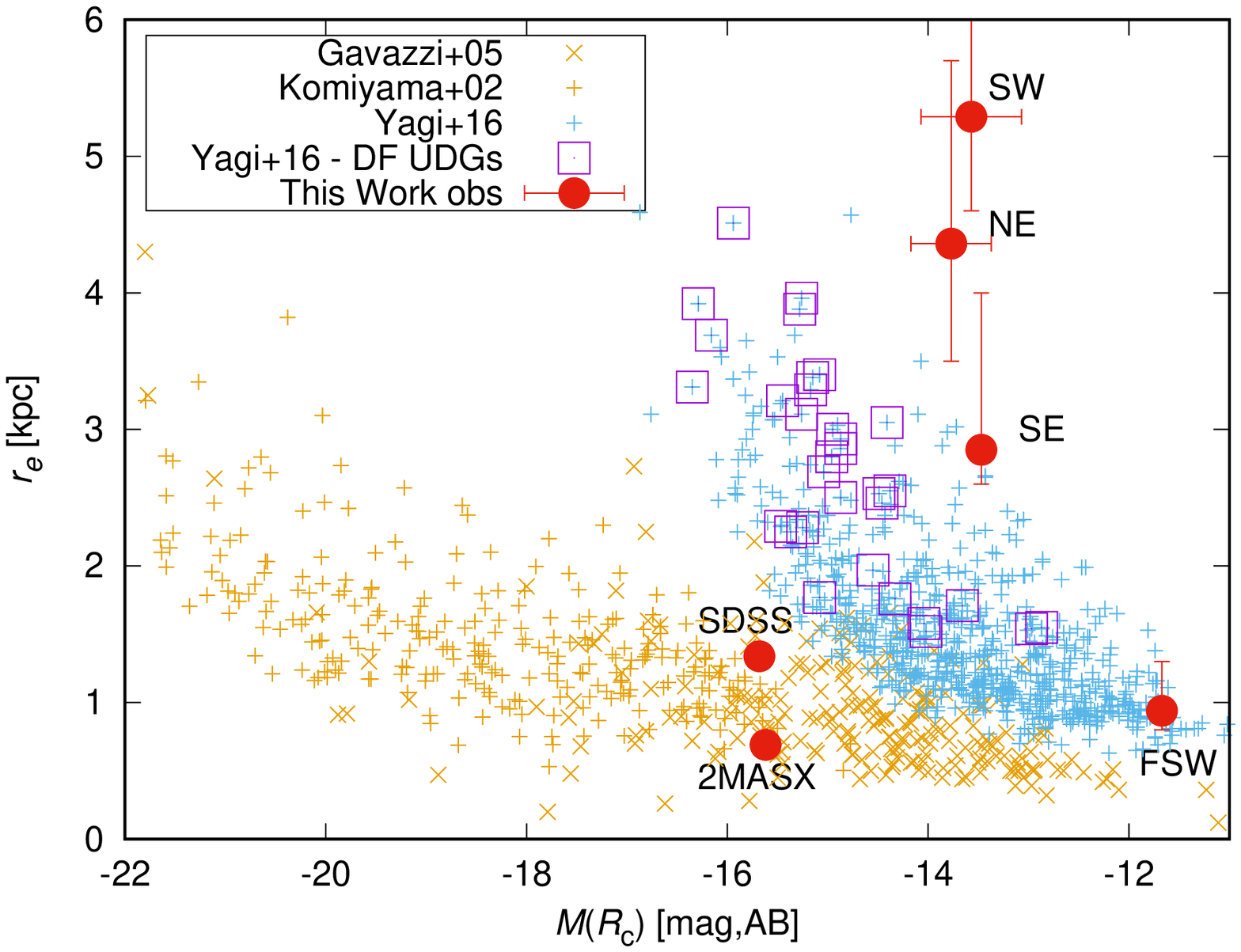}

\includegraphics[clip,width=6cm,angle=-90]{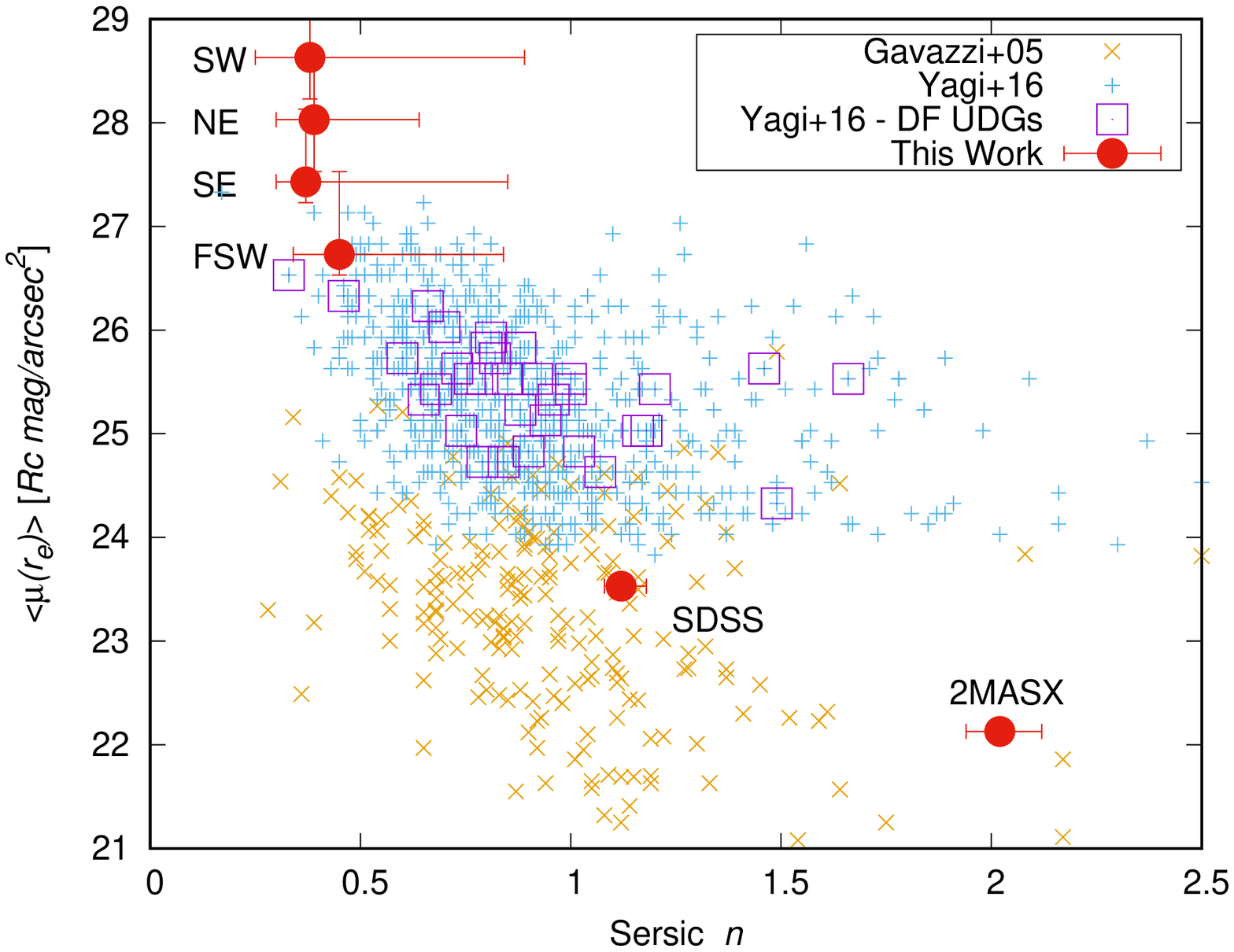}
\end{center}
\caption{(\textit{Top}) Comparison of the objects found around the NGC\,1068 with some similar objects in the literature in the absolute magnitude versus effective radius diagram. (\textit{Bottom}) The same figure for the S{\'e}rsic index versus $\langle\mu(r_\mathrm{e})\rangle$ (average surface brightness within the effective radius). Note that FSW, SDSS, and 2MASX are LSB-FSW, SDSS\,J0242, and 2MASX\,J0242, respectively.}
\label{fig7}
\end{figure}

We further estimate the total magnitude of the loop. First, we assume the geometry of the loop is a torus whose major and minor radii are 74.6 kpc and 3.5 kpc, respectively. The projected area of the loop is 3.5$\times 10^{5}$ arcsec$^{2}$.
Since some part of the loop is not seen even in our deep data, we roughly assume that the mean surface brightness of the loop would be $\sim$29 mag arcsec$^{-2}$. Then, the total apparent magnitude of the loop is about 15 mag, and the absolute magnitude is about $-16$ mag. Though the error might be large, the absolute magnitude is comparable to those of normal dwarf galaxies, and the scenario would be valid that the loop is a remnant of an infalled satellite.
The above absolute magnitude corresponds to the luminosity of $L \sim 2\times10^{8} L_{\solar}$, where we assume that the absolute magnitude of the Sun in r-band is 4.64 \citep{Blanton2007}.
Given the mass-to-luminosity ratio, $M/L \sim 10$ \citep{RH94}, we obtain the mass of $M \sim 2\times10^{9} M_{\solar}$. Since the nucleated fraction in galaxies is higher for larger-mass galaxies \citep{Miller15}, it is expected that the infalled satellite would have a supermassive black hole (hereafter SMBH) with the mass $M_{\rm SMBH} \sim 2\times10^{6} M_{\solar}$, given the stellar to SMBH mass ratio of 1000 (e.g., \cite{KH2013}). Here, it is reminded that one of the satellite galaxies of M\,31, M\,32, has the stellar mass $M \sim 3\times10^{9} M_{\solar}$ and $M_{\rm SMBH} \sim 3\times10^{6} M_{\solar}$ (\cite{KH2013}). We may imagine that M\,32 will merge into M\,31 and then we will see a similar loop structure around M\,31 several Gyrs after.

\begin{figure}
 \begin{center}
  \includegraphics[width=8cm]{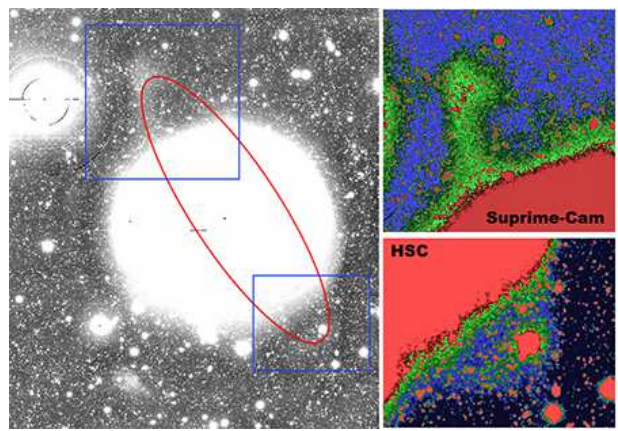} 
 \end{center}
\caption{A possible loop structure which connects UDO-NE and UDO-SW. A contrast-enhanced image for UDO-NE (upper right) is from the Suprime-Cam data, while that for UDO-SW is from the HSC images.}\label{fig5}
\end{figure}

Here, let us re-visit UDO-SE. Although it may be a tidal dwarf galaxy, we cannot exclude a possibility that it is an isolated faint dwarf galaxy. If this is the case, we can set constraints on its mass. UDO-SE lies at projected distance of 43.9 kpc. Assuming that the total mass of NGC\,1068 is $\sim 3\times 10^{11} M_{\solar}$ within the 43.9 kpc sphere, and the self gravity of UDO-SE maintains its shape at least in $<2$ kpc, the mass of UDO-SE in 2 kpc radius should be larger than $8.5 \times 10^{7} M_{\solar}$\citep{binneytremine87}. As the luminosity of UDO-SE is $\sim 1.6 \times 10^{7} L_{\solar}$ in $r$ band, the $M/L$ of UDO-SE is $\gtrsim 5.3$, which is similar to the nearby normal dwarf and the LSB galaxies (average $\sim 7.7$; \cite{swaters11}). Therefore, UDO-SE is probably not a simple tidal debris without dark matter, but seems to be a normal dwarf galaxy with its own dark-matter halo. The stretched shape of the UDO-SE, however, suggests that it is currently in a disrupting phase. If this is the case, the galaxy passed near NGC\,1068 (hence stronger tidal force) in the recent past. Or, the $M/L$ ratio could actually be smaller than the normal dwarfs. If the close encounter would have stimulated the star-formation on UDO-SE, the lower $M/L$ may be expected. A future color information for this object is necessary to fix this issue.

\subsection{Possible origins of the loop, UDOs, and warped disk}

As we have already discussed, UDO-SW and UDO-NE may be a part of the loop. 
The most likely origin of this loop is that the stellar envelope of a satellite dwarf galaxy has been stripped by the interaction with NGC\,1068. The best example of such a tidally liberated loop is seen around the edge-on spiral galaxy NGC\,5907 \citep{martinez-delgado08}. We show the possible interpretation of the single, circular loop in figure~\ref{fig5}. However, the actual loop structure may be more complex as seen in the case of NGC\,5907. In this respect, we cannot exclude a possibility that UDO-SE is a remnant that was made during the process of the loop formation.

If there was the past interaction event that has led the formation of the loop, one may consider that the ``Banana'' structure found at the outer disk (section 3.3) could also be the fossil record of the event. Indeed, the appearance of the NGC\,1068 outer disk is strikingly similar to the result of a minor-merger simulation presented in \citet{kazantzidis08}. They claimed that the duration of such outer one-arm or ripple structures are not short-lived but could last as long as $\sim$ 4 Gyrs after the minor merger. A similar simulation was also made by \citet{purcell11}. They explained the origin of a ring or ripple structure on a smooth disk by the infall of a single satellite galaxy. It is interesting to speculate that the core of the satellite galaxy may have sunk into the center of NGC\,1068 subsequently and have turned on the AGN activity as we see today \citep{Tani99}. We will discuss about this in the next section.

\subsection{Comments on a nucleated minor merger for the triggering AGN}

NGC\,1068 is an ordinary looking spiral galaxy although it belongs to the M\,77 (= NGC\,1068) group of galaxies together with NGC\,1055, and so on \citep{HG82}. Its morphology is classified as (R)SA(rs)b \citep{RC3}. Hence, it is generally thought that NGC\,1068 can be regarded as an isolated galaxy that experienced no past minor merger. However, several lines of evidence for a past minor merger in NGC\,1068 have been accumulated to date.

\subsubsection{Supporting evidence for a minor merger}

(1) Kinematically decoupled two stellar systems in the nuclear region:  
\citet{GL97} made spectroscopic investigation of the central 20 arcsec $\times$ 40 arcsec region of NGC\,1068 using the Ca\emissiontype{II} triplet line, and identified the off-centering of the rotation axis between the inner ($r <$ 3 pc) and the outer ($r > $ 5 pc) regions. This result indicates the presence of non-symmetric gravitational potential in the nuclear region. The most plausible interpretation is that this non-symmetric property is attributed to a past minor merger.
It is here reminded that the satellite galaxy should have a nucleus (i.e., an SMBH) in its center. Otherwise, the minor merger cannot affect the nuclear stellar dynamics in its host galaxy (see 4.3.2).

(2) The narrow line region inclined from the rotational axis of the main disk: 
The narrow line region (NLR) is an ionized gas-cloud system with a scale of 100 pc to 1 kpc, and is one of important characteristics of AGNs. The NLR is photoionized by the central engine although shock heating also works to some extent. Since the central engine is surrounded by a molecular/dusty torus with a typical radius is 0.1 pc to 10 pc (e.g.,\cite{TM98}), ionizing photons from the central engine are generally collimated by the torus. In fact, it is often observed that the NLR shows a biconical structure \citep{SK96}. 
This property is also observed in NGC\,1068. If the gas accretion onto the SMBH occurs taking an orbit in the plane of the main disk of NGC\,1068, the torus would locate in the main disk plane, and thus the biconical NLR should be observed along the rotational axis of the main disk. However, the NLR of NGC\,1068 is highly inclined from the rotational axis ($\sim$ 50 degree; \cite{C02}). This means that the torus plane itself is inclined from the main disk. We note that such a property is often observed in a number of Seyfert galaxies, in particular, type-2 Seyferts \citep{SK96}. This random nature of NLRs in Seyferts favors the minor-merger driven triggering of AGNs \citep{Tani99,Tani13}.

(3) The edge-on view molecular/dusty torus in the central 0.1 pc region:
The molecular/dusty torus in NGC\,1068 has been probed by using the water maser emission at 22 GHz (\cite{G96_H2O}, 2001, 2004; \cite{Greenhill96}); its inner and outer radii are 0.4 pc and 0.7 pc, respectively \citep{Greenhill96}. The observed rotation curve of H$_{2}$O maser emission provides a nuclear mass within a radius of 0.7 pc, $\sim 1 \times 10^7 ~ M_{\solar}$ at the distance of NGC\,1068, 15 Mpc \citep{Greenhill96}. Based on this observation, \citet{lodato03} obtained the mass of the central SMBH as $8 \times 10^{6} M_{\solar}$ at the distance of NGC\,1068, 14.4 Mpc. At the distance adopted in this paper, 15.9 Mpc, the nuclear mass within a radius of 0.7 pc, $\sim 1 \times 10^{7} M_{\solar}$. Since the mass of the putative secondary SMBH supplied by the merging satellite galaxy is a few $\times 10^{6} M_{\solar}$ at most, its contribution to the nuclear mass is acceptably small.

The observed torus is the edge-on view while the main disk of NGC\,1068 is not viewed as so; the optical axial ratio of NGC\,1068, $b/a$ = 0.89 gives a viewing angle $i = 63^{\circ}$, being far from the edge-on view ($a$ = 540.0 arcsec and $b$ = 480.6 arcsec are given in the NED). 
The edge-on view nature of the torus explains why the central engine and the broad line region cannot be seen because they are hidden by the optically-thick dusty torus while they can be seen in the polarized light through the electron scattering \citep{AM85}. Here, an important point is that the torus is kinematically decoupled from the main disk. If the material of the torus could be supplied from accreting gas clouds in the nuclear disk region of NGC\,1068, the torus would be located in the main disk plane. On the other hand, the minor merger scenario naturally explains this kinematically decoupled property because minor mergers occur by taking a randomly oriented orbit rather than the coplanar orbit.

\subsubsection{On the importance of a nucleated minor merger}

Minor mergers can be classified into the following two types.
One is nucleated minor mergers in which a nucleated satellite galaxy merges into its host galaxy. The other one is non-nucleated minor mergers in which a non-nucleated satellite galaxy merges into its host galaxy. Here, the term of ``nucleated'' means that a satellite galaxy has a nucleus (a SMBH). In the case of non-nucleated minor mergers, stars in a satellite galaxy are scattered into the galactic disk of its host galaxy during the course of the minor merging process (e.g., \cite{Khan12}). On the other hand, in the case of nucleated minor mergers, its nucleus (a SMBH) can migrate into the nuclear region of the host galaxy while most stars in the satellite galaxy are also scattered into the host disk.
Therefore, this type of minor mergers can trigger the nuclear activity in the host nucleus because the dynamical effect caused by the sinking secondary nucleus leads to the efficient gas fueling onto the primary nucleus. It is noted that 
intense circumnuclear ($r \sim 100$ pc) star formation is also triggered by the dynamical effect of the secondary nucleus before going into the very nuclear region \citep{TW96}.

Here, let us go back to the case of NGC\,1068. It has been recognized that NGC\,1068 has a luminous circumnuclear starburst (CNSB) region that shows a ring-like morphology at a radius of $r \sim$ 1 kpc with a luminosity of $L_\mathrm{CNSB}\sim 10^{11} L_{\solar}$ (\cite{T84}; see also \cite{YT97}). If NGC\,1068 is an isolated galaxy for a long time, it seems difficult to explain what mechanism causes the luminous CNSB as well as the AGN phenomenon. However, if NGC\,1068 would experience a nucleated minor merger, both its CNSB and AGN phenomenon could be explained without taking account of any other physical processes because the sinking secondary SMBH causes a luminous CNSB first and then gas fueling onto the primary SMBH \citep{YT97, Tani13}. Namely, during the course of a nucleated minor merger, a CNSB comes first, and then an AGN phenomenon comes later. This sequence is also proposed for major mergers \citep{H08}. It is also noted that starburst galaxies tend to have more disturbed morphology than Seyfert galaxies \citep{Schawinski10}, suggesting again that a circumnuclear starburst comes first and then AGN comes later.

Then, let us consider which is the case: i.e., a nucleated minor merger or not in the case of NGC\,1068. First, we revisit the molecular/dusty torus probed by the water maser emission. In addition to NGC\,1068, the best-studied torus is that in NGC\,4258 \citep{Miyoshi95}. However, there is a significant difference in the dynamical properties of the tori between NGC\,4258 and NGC\,1068. Namely, the torus of NGC\,4258 shows the Keplerian rotation while that of NGC\,1068 shows non-Keplerian rotation (the velocity $v$ is proportional to $r^{-1/3}$; \cite{Greenhill96}). \citet{MT97} suggested that the non-Keplerian rotation in NGC\,1068 can be attributed to the dynamical effect of a bar structure inside the torus. However, it is noted that the dynamical effect of a bar is equivalent to that of a pair of SMBHs; one is the SMBH already resided in the center of NGC\,1068 while the other one is a SMBH supplied by a sinking satellite galaxy.

Another important observational evidence for the nucleated minor merger in NGC\,1068 is the presence of the kinematically decoupled two stellar system in the nuclear region reported by \citet{GL97}. The two off-centered stellar system between the inner ($r <$ 3 pc) and the outer ($r > $ 5 pc) regions cannot be attributed to a non-nucleated minor merger because such a merger can never affect the stellar dynamics in the very nuclear region. The nucleated minor merger hypothesis also explains two distinct, episodic star formation events: $\sim$ 300 Myr ago in the central 180 pc (in radius) region, and $\sim$ 30 Myr ago in the ring at $r \simeq$ 100 pc region \citep{SB12}.

Here, we propose a hypothesis of the past history of NGC\,1068.
Several Gyrs ago, a nucleated satellite galaxy began to fall onto NGC\,1068.
The tidal force of NGC\,1068 stripped stars of the satellite galaxy to create 
the observed loop structure. After a few orbital rotations, the satellite galaxy gradually falls onto near nuclear regions. Recently, 500 Myrs ago, the SMBH of the satellite galaxy approached to a few hundreds pc from the SMBH of NGC\,1068 and ignited the central starburst; e.g., \citet{YT97}. Then, the SMBH continues to sink into the nuclear region (i.e., $<$ 1 pc), and finally stimulates the AGN by fueling the gas of the NGC\,1068 disk.
This hypothesis would be examined by future observations. The color of the UDOs would set constraints on their stellar population. Also, spectroscopic observations of the Banana, the Ripple, and the UDOs will give us kinematic information of the structures and enables us to reconstruct the whole merging history.

\section{Conclusion}
Minor mergers have been considered as one of triggering mechanisms of AGNs,
in particular, Seyfert galaxies \citep{Tani99,Tani13}.
Therefore, it is important to search for firm observational evidence for minor mergers in Seyfert galaxies. However, since tidal effects of minor mergers are generally weak, it has been difficult to identify convincing evidence for minor mergers in Seyfert galaxies. 

In this paper, we have presented our deep optical imaging of the archetypical type 2 Seyfert galaxy, NGC\,1068 (M\,77) carried out by using both Hyper Suprime-Cam and Suprime-Cam on the 8.2 m Subaru Telescope where the Suprime-Cam data were taken from the SMOKA data archive. Our deep optical data has enabled us to do the census of very faint features around NGC\,1068 at $\sim 29$ mag arcsec$^{-2}$ surface brightness level.

The main results and conclusion are as follows.
\begin{enumerate}
\item It is found that the three ultra defuse objects (UDOs) appear to be associated with NGC\,1068; they are called as UDO-NE, UDO-SW, and UDO-SE. Note that UDO-SE was previously identified in \citet{bakos12} based on the SDSS Stripe 82 data.

\item The three UDOs around NGC\,1068 have the following properties; i) their S{\'e}rsic indices are fairly small ($< 0.5$), and ii) their average effective surface brightness is very low ($\langle\mu(r_\mathrm{e})\rangle > 26$ mag arcsec$^{-2}$). Therefore, it is suggested that they are different populations from UDGs found in the cluster environment. A plausible idea is that they are tidally induced objects during the course of a past minor merger.

\item The detailed morphological properties of both UDO-NE and UDO-SW suggest that these two structures would make a single loop surrounding the main body of NGC\,1068, being similar to those found in our Galaxy \citep{newberg02}, NGC\,5907\citep{martinez-delgado08}, and so on. If this is the case, the two UDOs (UDO-NE and UDO-SW) is a so-called stream structure that is a tidal remnant of a merging satellite galaxy.

\item Another UDO-like object is also found at far SW. This object is $\sim2$ magnitudes fainter and smaller by a factor of 3--5 than those of the three UDOs. This object may belong to a class of low surface brightness galaxy and thus we call it LSB-FSW. Since there is no redshift information, it is difficult to confirm its physical connection to NGC\,1068. However, it is interesting to note that this object is located at the line connecting both UDO-NE and UDO-SW. Therefore, we suggest that this object may be a relic of the formation of the loop structure.

\item In addition to the UDOs, it is also found that an outer one-arm structure is emanated from the SW edge of the main body of NGC\,1068; we call this the Banana structure. And also, a ripple-like structure is found at the opposite side of the outer disk. A plausible idea is again that they are tidally-induced structures of a past minor merger.

\item All the above observational results provide evidence for a past minor merger occurred in NGC\,1068 several billions years ago. 
\end{enumerate}

Since NGC\,1068 appears to be an ordinary looking, beautiful and symmetric spiral galaxy, we tend to regard this galaxy as an isolated galaxy. Therefore, it has been often considered that the nuclear activity in NGC\,1068 is not rated to any merger events (\cite{gb2014}, and references therein). Yet, several lines of supporting evidence for a minor merger in NGC\,1068; (a) the presence of kinematically decoupled two stellar population systems within the central several parsec region, (b) the presence of the molecular-dusty torus whose rotational axis is almost orthogonal to the main disk of NGC\,1068, and (c) the nuclear radio jet axis as well as that of NLR is also highly inclined with respect to the rotational axis of the main disk.
Now, together with our new discovery presented in this paper, all these properties can be reasonably interpreted as the effect of a past minor merger. Another important aspect is that a past minor merger surely affected the dynamics 
of central parsec region. This strongly suggests that the putative satellite galaxy merged into NGC\,1068 is a nucleated satellite galaxy such as M\,32 which has an SMBH with mass of $\sim10^{6}$ M$_{\solar}$ (\cite{KH2013}, and references therein).

\begin{ack}
We greatly appreciate the anonymous referee for his/her useful comments and suggestions that helped us to improve our paper.
We thank Yutaka Komiyama for providing us their unpublished $r_\mathrm{e}$ catalog for Coma Dwarfs as well as his useful comments. We also thank Mikito Tanaka, and Keiichi Wada for their useful comments. We thank Hisanori Furusawa, Michitaro Koike, and the member of the HSC pipeline HelpDesk for their help for our HSC data analysis. IT thanks Fumiaki Nakata for his efforts on building the HSC data analysis environment in Subaru Observatory. We appreciate the operation staff of the Subaru Observatory for the execution of our Queue program. We would also like to thank Michael R. Blanton for providing us the SDSS color image of NGC\,1068 for our figure~\ref{fig1}.
This work has made use of the SDSS database and the NASA/IPAC Extragalactic Database (NED). We also used the photometric and astrometric data from the Pan-STARRS1 Survey (PS1) database during the HSC data analysis. This work has also made use of the SMOKA archive and the computer systems at Astronomical Data Analysis Center of National Astronomical Observatory of Japan (NAOJ). This work was supported by JSPS KAKENHI Grant Number 16H02166 (YT).
Finally, we wish to recognize and acknowledge the very significant cultural role and reverence that the summit of Maunakea has always had within the indigenous Hawaiian community. We are most fortunate to have the opportunity to conduct observations from this mountain.
\end{ack}

\appendix
\section*{Possible effects of the CCD edges on the shape of the UDOs}
There are some concerns about whether the shape of the UDO-SW and NE are affected by the CCD edges, because the left side of UDO-NE and right side of UDO-SW that makes up the loop are apparently aligned with the CCD columns. To see whether the morphology is affected by the CCD edges, we compare the exposure map with the morphology of the UDO-SW and NE. 

We first divide 33 exposure dataset into 3 subsets, each of which contains 11 frames. We then combine the data in each subset. 
We do not use the last 2 of the whole 35 frames for the test, since it is better to keep the number of each subset the same.
We also generate the exposure map for each. If the morphology of the UDOs are significantly different among the subsets, or if the exposure map pattern and the outer edges of the UDO-NE and/or SW show a clear correlation, it indicates that the morphology is affected by the CCD columns. 

The result is shown in figure~\ref{fig_a1}. We can see that the CCD gaps as seen in the exposure map sometimes show rectilinear patterns either on the sky or on the stellar halo.  However, the overall morphology of three UDOs in each subset appears to be the same as the total image in which all the 33 exposures are combined. Thus we conclude that the CCD edges does not significantly affect the overall morphology of the UDO-SW and NE. 
We should note that the shape of objects in the frames combined the 33 frames and all 35 frames are indistinguishable.

We also compare the morphology of each UDOs in HSC, Suprime-Cam, and the SDSS DR12 data. The result is shown in figure~\ref{fig_a2}. Overall, the shapes of objects in the HSC data are fairly similar to those in the Suprime-Cam data. Although the SDSS data is shallow and largely affected by worse sky subtraction, we can see the trace of the three UDOs. Thus we conclude that their presence is real.

\begin{figure*}
 \begin{center}
 \includegraphics[width=13cm]{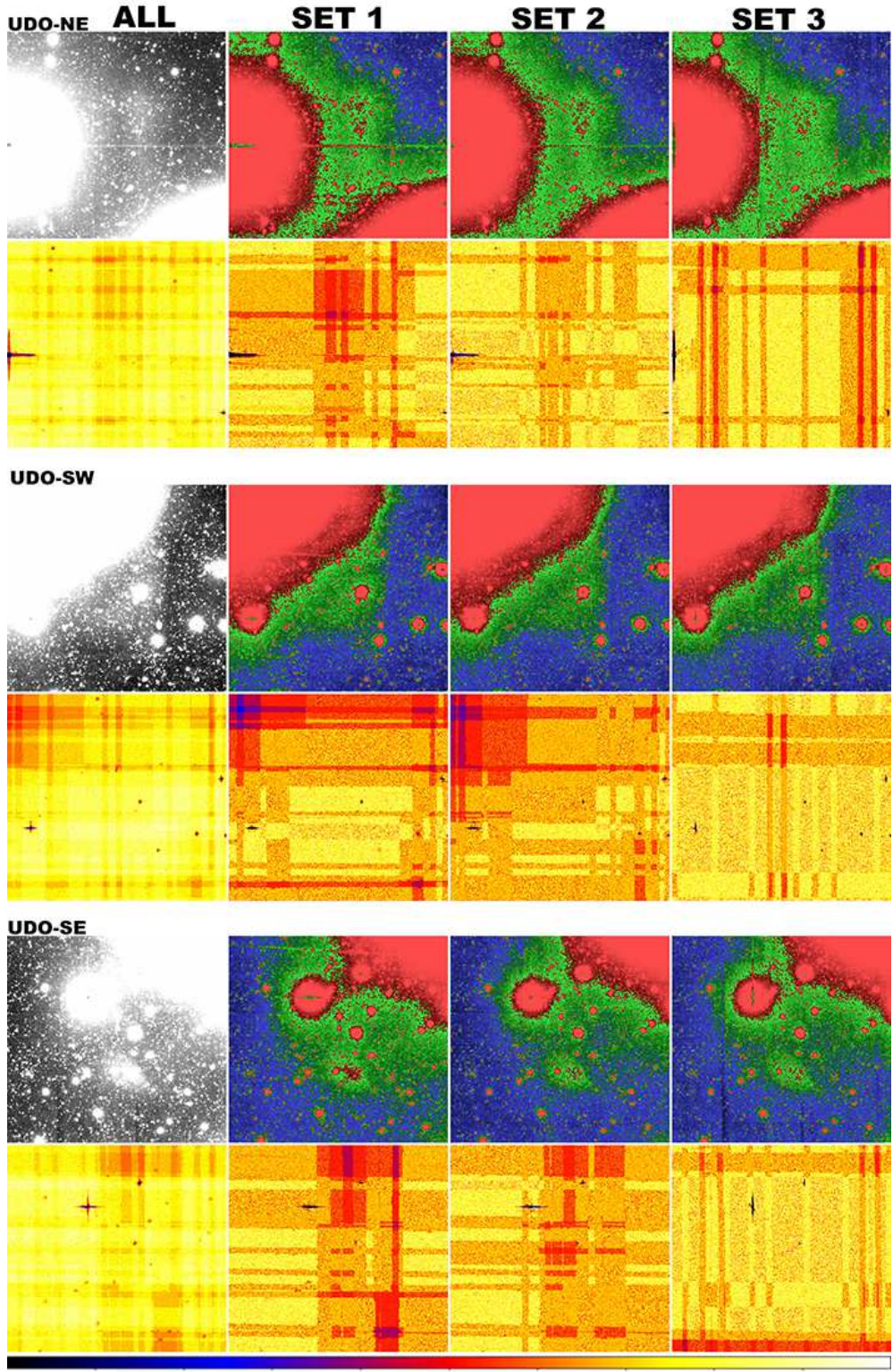} 
 \end{center}
\caption{Comparison of the UDO shape among the subsamples. Each set of 8 tiled images at the top, middle, and bottom shows the result for UDO-SE, UDO-NE, and UDO-SW, respectively. In each grouped images, the top left image (monochrome) shows the image with all the 33 frames combined. The next three images are the result for the subsets, each of which is the average of the 11 frames. At the bottom of each panel, we show the exposure map for each combined subset, with the scale bar at the bottom of the figure (left$= 0 \%$, right$= 100\%$ of the total exposures for each subset).
}\label{fig_a1}
\end{figure*}

\begin{figure*}
 \begin{center}
 \includegraphics[width=13cm]{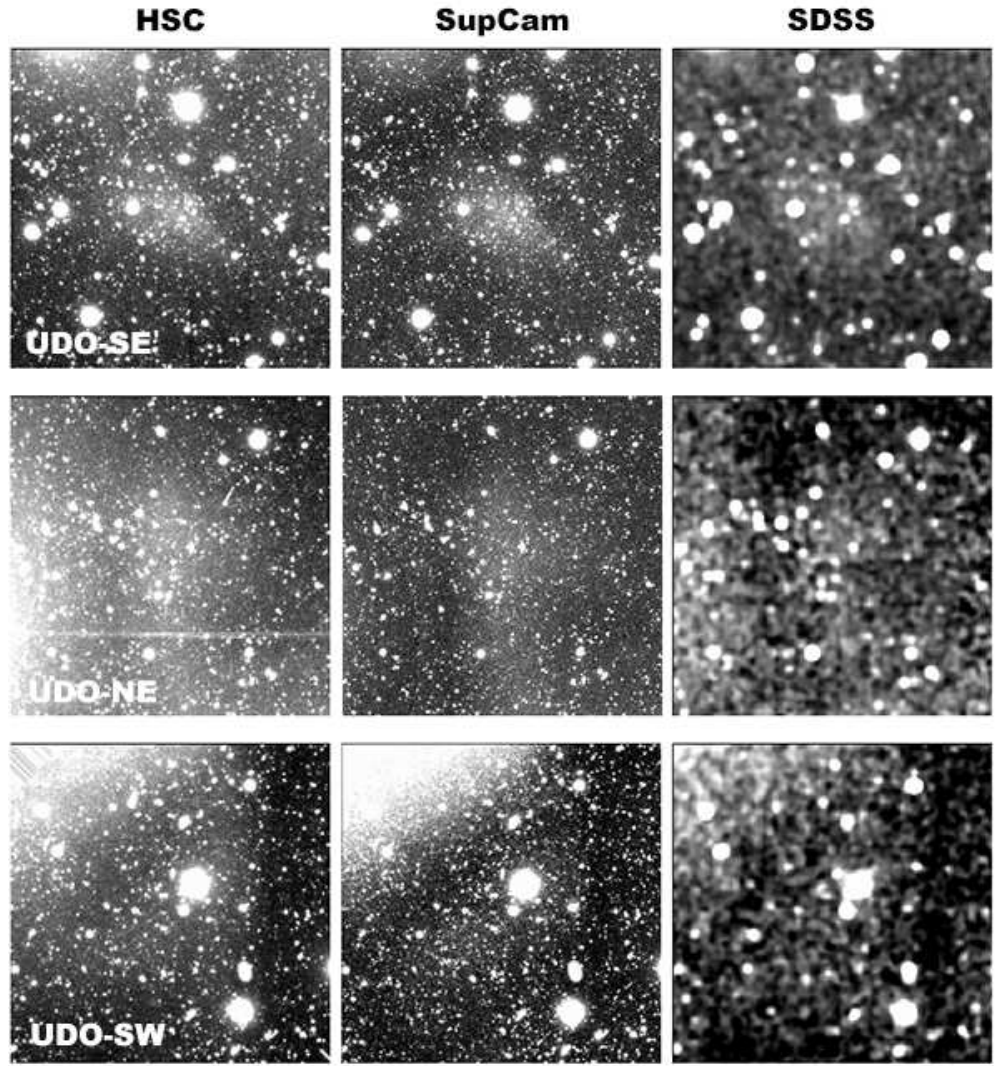} 
 \end{center}
\caption{Comparison of the appearance of the UDOs on the images taken by the HSC (left), the Suprime-Cam (middle), and the SDSS DR12 (right: here we use the $r$-band data). Note that the SDSS images are after applying the smoothing ($3 \times 3$ pixels block average $+$ the Gaussian smoothing with $\sigma=2$ pixels) to show the faintest features. Also note that the SDSS data are largely affected by the poor sky subtraction, and thus any morphological information from the data may not be reliable. However, we can barely see a hint of these UDOs in the images.
}\label{fig_a2}
\end{figure*}


\begin{thebibliography}{}
\bibitem[Abazajian et al.(2009)]{SDSS-DR7} Abazajian, K.~N., Adelman-McCarthy, J.~K., Ag\"ueros, M.~A. et al.\ 2009,\apjs, 182, 543
\bibitem[Alam et al.(2012)]{SDSS-DR12} Alam, S, Albareti, F.~D, Allende Prieto, C. et al.\ 2015, \apjs, 219, 12
\bibitem[Antonucci \& Miller(1985)]{AM85} Antonucci, R. R. J., \& Miller, J. S.\ 1985, \apj, 297, 621
\bibitem[Baba et al.(2002)]{Baba2002}
Baba, H., Yasuda, N., Ichikawa, S.-I. et al. 2002, in ASP Conf. Ser. 281: ADASS XI, ed. Bohlender, D.A., Durand, D. \& Handley, T.H., 298
\bibitem[Bakos \& Trujillo(2012)]{bakos12} Bakos, J., \& Trujillo, I.\ 2012, arXiv:1204.3082 
\bibitem[Baldwin et al.(1981)]{Baldwin1981} Baldwin, J.~A., Phillips, M.~M., \& Terlevich, R. 1981, \pasp, 93, 5
\bibitem[Barnes \& Hernquist(1992)]{BH92} Barnes, J.~E., \& Hernquist, L.\ 1992, \nat, 360, 715 
\bibitem[Bertin \& Arnouts(1996)]{bertin} Bertin, E., \& Arnouts, S.\ 1996, \aaps, 117, 393
\bibitem[Binney \& Tremine(1987)]{binneytremine87} Binney, J., \& Tremine, S. \ 1987, Galactic Dynamics (Princeton University Press), chap. 7 
\bibitem[Blanton \& Roweis(2007)]{Blanton2007} Blanton, M.~R., \& Roweis, S. \ 2007, \aj, 133, 734 
\bibitem[Bosch et al.(2017)]{bosch17} Bosch, J., Armstrong, R., Bickerton, S., et al.\ 2017, \pasj, in press (arXiv:1705.06766)
\bibitem[Cecil et al.(2002)]{C02} Cecil, G., Dopita, M. A., Groves, B., WIson, A. S., Ferruit, P., Pecontal, E.,  \& Binette, L. \ 2002, \apj, 568,  627
\bibitem[Coleman et al.(1980)]{cww80} Coleman, G.~D., Wu, C.-C., \& Weedman, D.~W.\ 1980, \apjs, 43, 393
\bibitem[Chambers et al.(2016)]{chambers16} Chambers, K.~C., Magnier, E.~A., Metcalfe, N., et al.\ 2016, arXiv:1612.05560 
\bibitem[Chonis et al.(2011)]{chonis11} Chonis, T.~S., Mart{\'{\i}}nez-Delgado, D., Gabany, R.~J., et al.\ 2011, \aj, 142, 166 
\bibitem[de Vaucouleurs et al.(1991)]{RC3} de Vaucouleurs, G., et al.\ 1991, Third  Reference Catalog of Bright Galaxies (RC3)
\bibitem[Dickel \& Rood(1978)]{dickel78} Dickel, J.~R., \& Rood, H.~J.\ 1978, \apj, 223, 391 
%
\bibitem[Duc et al.(2014)]{duc2014} Duc, P.-A., Paudel, S., McDemid, R.~M., et al.\ 2014, \mnras, 440, 1458
\bibitem[Duc et al.(2015)]{duc2015} Duc, P.-A., Cuillandre, J.-C., Karabal, E., et al.\ 2015, \mnras, 446, 120
%
\bibitem[Fukugita et al.(1995)]{fukugita95} Fukugita, M., Shimasaku, K., \& Ichikawa, T.\ 1995, \pasp, 107, 945 
\bibitem[Fukugita et al.(2007)]{F07} Fukugita, M., Nakamura, O., Okamura, S., et al.\ 2007, \aj, 134, 579
\bibitem[Furuya \& Taniguchi(2016)]{furu_tani16} Furuya, R.~S., \& Taniguchi, Y.\ 2016, \pasj, 68, 103 
\bibitem[Gavazzi et al.(2005)]{gavazzi05} Gavazzi, G., Donati, A., Cucciati, O., et al.\ 2005, \aap, 430, 411
\bibitem[Graham \& Guzm{\'a}n(2003)]{graham03} Graham, A.~W., \& Guzm{\'a}n, R.\ 2003, \aj, 125, 2936 
\bibitem[Graham \& Driver(2005)]{grahamdriver05} Graham, A.~W., \& Driver, S.~P.\ 2005, PASA, 22, 118 
\bibitem[Garc{\'{\i}}a-Burillo et al.(2014)]{gb2014} Garc{\'{\i}}a-Burillo, S., Combes, F., Usero, A., et al.\ 2014, \aap, 567, A125 
\bibitem[Garc{\'{\i}}a-Lorenzo et al.(1997)]{GL97} Garc{\'{\i}}a-Lorenzo, B., Mediavilla, E., Arribas, S., \& del Burgo, C. \ 1997, \apj, 483, L99
\bibitem[Gallimore et al.(1996)]{G96_H2O} Gallimore, J.~F., Baum, S.~A., O'Dea, C.~P., Brinks, E., \& Pedlar, A.\ 1996, \apj, 462, 740 
\bibitem[Gallimore et al.(2001)]{G01} Gallimore, J.~F., Henkel, C., Baum, S.~A., et al.\ 2001, \apj, 556, 694 
\bibitem[Gallimore et al.(2004)]{G04} Gallimore, J.~F., Baum, S.~A., \& O'Dea, C.~P.\ 2004, \apj, 613, 794 
\bibitem[Goto(2005)]{Goto2005} Goto, T. 2005, \mnras, 357, 937
\bibitem[Greenhill et al.(1996)]{Greenhill96} Greenhill, L.~J., Gwinn, C.~R., Antonucci, R., \& Barvainis, R.\ 1996, \apjl, 472, L21 
\bibitem[H{\"a}ussler et al.(2007)]{haussler07} H{\"a}ussler, B., McIntosh, D.~H., Barden, M., et al.\ 2007, \apjs, 172, 615
\bibitem[Holwerda et al.(2014)]{holwerda14} Holwerda, B.~W., Mu{\~n}oz-Mateos, J.-C., Comer{\'o}n, S., et al.\ 2014, \apj, 781, 12 
\bibitem[Hopkins et al.(2008)]{H08} Hopkins, P. F., Hernquist, L. C., Cox, T. J.,  \&  Keres, D.\ 2008, \apjs, 175, 356 
\bibitem[Huchra \& Geller (1982)]{HG82} Huchra, J. P., \& Geller, M. J.\ 1982, \apj, 252, 437
\bibitem[Ji et al.(2014)]{ji14} Ji, I., Peirani, S., \& Yi, S.~K.\ 2014, \aap, 566, A97 
\bibitem[Kazantzidis et al.(2008)]{kazantzidis08} Kazantzidis, S., Bullock, J.~S., Zentner, A.~R., Kravtsov, A.~V., \& Moustakas, L.~A.\ 2008, \apj, 688, 254-276
\bibitem[Kauffmann et al.(2003)]{Kauffmann2003} Kauffmann, G., Heckman, T.~M., Tremonti, C., et al. 2003, \mnras, 346, 1055
\bibitem[Khan et al.(2012)]{Khan12} Khan, F.~M., Preto, M., Berczik, P., et al.\ 2012, \apj, 749, 147 
\bibitem[Kewley et al.(2001)]{Kewley2001} Kewley, L.~J., Dopita, M.~A., Sutherland, R.~S., Heisler, C.~A., Trevena, J. 2001, \apj, 556, 121
\bibitem[Koda et al.(2015)]{koda15} Koda, J., Yagi, M., Yamanoi, H., \& Komiyama, Y.\ 2015, \apjl, 807, L2 
\bibitem[Komiyama et al.(2002)]{komiyama02} Komiyama, Y., Sekiguchi, M., Kashikawa, N., et al.\ 2002, \apjs, 138, 265 
\bibitem[Kormendy \& Ho(2013)]{KH2013} Kormendy, J., \& Ho, L.~C.\ 2013, \araa, 51, 511
\bibitem[Lang et al.(2010)]{Lang2010}
Lang, D., Hogg, D. W., Mierle, K., Blanton, M., Roweis, S.\ 2010, \aj, 139, 1782\bibitem[Malkan et al.(1998)]{malkan98} Malkan, M.~A., Gorjian, V., \& Tam, R.\ 1998, \apjs, 117, 25
\bibitem[Lodato \& Bertin(2003)]{lodato03} Lodato, G., \& Bertin, G.\ 2003, \aap, 398, 517
\bibitem[Mart{\'{\i}}nez-Delgado et al.(2008)]{martinez-delgado08} Mart{\'{\i}}nez-Delgado, D., Pe{\~n}arrubia, J., Gabany, R.~J., et al.\ 2008, \apj, 689, 184
\bibitem[Mart{\'{\i}}nez-Delgado et al.(2010)]{martinez-delgado10} Mart{\'{\i}}nez-Delgado, D., Gabany, R.~J., Crawford, K., et al.\ 2010, \aj, 140, 962 
\bibitem[Mart{\'{\i}}nez-Delgado et al.(2012)]{martinez-delgado12} Mart{\'{\i}}nez-Delgado, D., Romanowsky, A.~J., Gabany, R.~J., et al.\ 2012, \apjl, 748, L24 
\bibitem[Mart{\'{\i}}nez-Delgado et al.(2015)]{martinez-delgado15} Mart{\'{\i}}nez-Delgado, D., D'Onghia, E., Chonis, T.~S., et al.\ 2015, \aj, 150, 116 
\bibitem[Merritt et al.(2016)]{merritt16} Merritt, A., van Dokkum, P., Danieli, S., et al.\ 2016, \apj, 833, 168 
\bibitem[Miller et al.(2015))]{Miller15} Miller, B.,  Gallo, E., Henny, E., et al.\ 2015, \apj, 799, id.98 
\bibitem[Miskolczi et al.(2011)]{miskolczi11} Miskolczi, A., Bomans, D.~J., \& Dettmar, R.-J.\ 2011, \aap, 536, A66 
\bibitem[Miyazaki et al.(2002)]{Miyazaki2002} Miyazaki, S., Komiyama, Y., Sekiguchi, M., et al. 2002, \pasj, 54, 833
\bibitem[Miyazaki et al.(2012)]{Miyazaki12} Miyazaki, S., Komiyama, Y., Nakaya, H., et al.\ 2012, \procspie, 8446, 84460Z
\bibitem[Miyoshi et al.(1995)]{Miyoshi95} Miyoshi, M.,Moran, J., Herrnstein, J., Greenhill, L., Nakai, N., Diamond, P., \& Inoue, M,\ 1995, \nat, 373, 127 
\bibitem[Mobasher et al.(2001)]{mobasher01} Mobasher, B., Bridges, T.~J., Carter, D., et al.\ 2001, \apjs, 137, 279 
\bibitem[Murayama \& Taniguchi(1997)]{MT97} Murayama, T., \& Taniguchi, Y.\ 1997, \pasj, 49, L13 
\bibitem[Newberg et al.(2002)]{newberg02} Newberg, H.~J., Yanny, B., Rockosi, C., et al.\ 2002, \apj, 569, 245 
\bibitem[Peng et al.(2002)]{peng02}  Peng, C.~Y., Ho, L.~C., Impey, C.~D., \& Rix, H.-W.\ 2002, \aj, 124, 266 
\bibitem[Peng et al.(2010)]{peng10} Peng, C.~Y., Ho, L.~C., Impey, C.~D., \& Rix, H.-W.\ 2010, \aj, 139, 2097
\bibitem[Purcell et al.(2011)]{purcell11} Purcell, C.~W., Bullock, J.~S., Tollerud, E.~J., Rocha, M., \& Chakrabarti, S.\ 2011, \nat, 477, 301 
\bibitem[Robert \& Haynes(1994)]{RH94} Robert, M. S., \& Heynes, M. P.\ 1994, \araa, 32, 115
\bibitem[S{\'a}nchez-Portal et al.(2000)]{SanchezPortal00} S{\'a}nchez-Portal, M., D{\'{\i}}az, {\'A}.~I., Terlevich, R., et al.\ 2000, \mnras, 312, 2 
\bibitem[Sanders et al.(1988)]{S88} Sanders, D.~B., Soifer, B.~T., Elias, J.~H., et al.\ 1988, \apj, 325, 74 
\bibitem[Schawinski et al.(2010)]{Schawinski10} Schawinski, K., Dowlin, N., Thomas, D., Urry, C.~M., \& Edmondson, E.\ 2010, \apjl, 714, L108 
\bibitem[S{\'e}rsic(1968)]{sersic} S{\'e}rsic, J.~L.\ 1968, Atlas de galaxias australes, Cordoba, Argentina: Observatorio Astronomico
\bibitem[Seyfert(1943)]{Sey43} Seyfert, C. K.\ 1943, \apj, 97, 28 
\bibitem[Schmitt \& Kinney(1996)]{SK96} Schmitt, H.~R., \& Kinney, A.~L.\ 1996, \apj, 463, 498 
\bibitem[Smirnova et al.(2010)]{smirnova10} Smirnova, A.~A., Moiseev, A.~V., \& Afanasiev, V.~L.\ 2010, \mnras, 408, 400 
\bibitem[Storchi-Bergamnn et al.(2012)]{SB12} Storchi-Bergamnn, T., et al.\ 2012, \apj, 755, 87 
\bibitem[Swaters et al.(2011)]{swaters11} Swaters, R.~A., Sancisi, R., van Albada, T.~S., \& van der Hulst, J.~M.\ 2011, \apj, 729, 118 
\bibitem[Taniguchi(1997)]{YT97} Taniguchi, Y.\ 1997,  \apj, 487, L17
\bibitem[Taniguchi(1999)]{Tani99} Taniguchi, Y.\ 1999, \apj, 524, 65 
\bibitem[Taniguchi(2013)]{Tani13} Taniguchi, Y.\ 2013, Galaxy Mergers in an Evolving Universe, ed. Sun, W-H; Xu, C. K.; Scoville, N. Z.; \& Sanders, D. B. (San Francisco: ASP), 477, 265 
\bibitem[Taniguchi \& Murayama(1998)]{TM98} Taniguchi, Y., \& Murayama, T.\ 1998, \apjl, 501, L25 
\bibitem[Taniguchi \& Shioya(1998)]{TS98} Taniguchi, Y., \& Shioya, Y.\ 1998, \apj, 501. L167 
\bibitem[Taniguchi \& Wada(1996)]{TW96} Taniguchi, Y., \& Wada, K.\ 1996, \apj, 469, 581 
\bibitem[Telesco et al.(1984)]{T84} Telesco, C. M., et al.\ 1984, \apj, 282, 427
\bibitem[Urry \& Padovani(1995)]{UP95} Urry, C.~M., \& Padovani, P.\ 1995, \pasp, 107, 803 
\bibitem[van Dokkum et al.(2015a)]{vDm15} van Dokkum, P.~G., Abraham, R., Merritt, A., et al.\ 2015, \apjl, 798, L45 
\bibitem[van Dokkum et al.(2015b)]{vDm15b} van Dokkum, P.~G., Romanowsky, A.~J., Abraham, R., et al.\ 2015, \apjl, 804, L26
\bibitem[van Dokkum et al.(2016)]{vDm16} van Dokkum, P.~G., Abraham, R., Brodie, J., et al.\ 2016, \apjl, 828, L6 
\bibitem[Willett et al.(2013)]{willett13}  Willett, K.~W., Lintott, C.~J., Bamford, S.~P., et al.\ 2013, \mnras, 435, 2835 
\bibitem[Yagi et al.(2013)]{yagi2013} Yagi, M., Suzuki, N., Yamanoi, H. et al.\ 2013, \pasj, 65, 22 
\bibitem[Yagi et al.(2016)]{yagi2016} Yagi, M., Koda, J., Komiyama, Y.\& Yamanoi, H.\ 2016, \apjs, 225, 11 
\bibitem[Yagi et al.(2017)]{yagi2017} Yagi, M., Yoshida, M., Komiyama, H. et al.\ 2017, \apj, 839, 65 
\bibitem[Yoshida et al.(2016)]{Yoshida2016} Yoshida, M., Yagi, M., Ohyama, Y., et al.\ 2016, \apj, 820, 48 
\end{thebibliography}
\end{document}